\documentclass[twocolumn,apj,numberedappendix,twocolappendix,iop]{openjournal}
\usepackage[T1]{fontenc}
\usepackage{graphicx}	
\usepackage{amsmath}	
\usepackage{newtxtext,newtxmath}    
\usepackage[dvipsnames]{xcolor}  
\usepackage[colorlinks,linkcolor=blue,citecolor=blue,urlcolor=blue ]{hyperref}
\usepackage{amssymb}	
\usepackage{dsfont}
\usepackage{ulem}
\usepackage{soul}
\usepackage{enumitem}
\usepackage{orcidlink}
\usepackage{hyperref}

\usepackage{natbib}

\usepackage{fontawesome}
\usepackage{lineno}

\usepackage{macros_vdbosch} 

\defcitealias{Dattathri.etal.25b}{D25}

\begin{document}


\title{Dynamics in the Cores of Self-Interacting Dark Matter Halos:\\
Reduced Stalling and Accelerated Core Collapse}

\shorttitle{Dynamics in the Cores of SIDM Halos}
\shortauthors{van den Bosch and Dattathri}

\author{Frank C. van den Bosch\orcidlink{0000-0003-3236-2068}}   
\author{Shashank Dattathri\orcidlink{0000-0002-7941-1149}}

\affiliation{Department of Astronomy, Yale University, PO. Box 208101, New Haven, CT 06520-8101}

\email{frank.vandenbosch@yale.edu}

\label{firstpage}


\begin{abstract}
  Self-interacting dark matter (SIDM) is an intriguing alternative to the standard cold dark matter (CDM) paradigm, which predicts that dark matter halos typically have large, isothermal cores. Numerical simulations have shown that dynamical friction ceases to operate in cores of (roughly) constant density, a phenomenon known as core stalling. In addition, such cores often are unstable to a dipole instability that gives rise to dynamical buoyancy, causing massive central objects to move outward. Thus far, these manifestations of core dynamics have only been demonstrated in collisionless systems. In this paper, we use idealized $N$-body simulations to study whether core stalling and dynamical buoyancy operate in SIDM halos. We find that if the self-interactions are sufficiently strong, neither core stalling nor buoyancy are present, and a massive perturber will quickly sink all the way to the center of its host. In doing so, it gravitationally contracts the core, catalyzing a strongly accelerated core collapse. The reason why core dynamics are so different in SIDM halos is that self-interactions drive the halo's distribution function to a featureless exponential, removing any inflections or plateaus that are responsible for the dipole instability and core stalling in the case of CDM. We discuss implications of our finding for constraining the nature of dark matter by using observations of massive objects such as supermassive black holes (SMBHs), globular clusters, and nuclear star clusters in the central regions of galaxies.
\end{abstract}

\keywords{
Dark Matter --
Dynamical Friction --
Gravitational Instability --
Galaxy Dynamics --
Galaxy Nuclei --
Galaxy Dark Matter Halos --
N-body simulations}

\maketitle


\section{Introduction}

Cores, which are regions of roughly uniform density, seem to be ubiquitous in the central regions of a wide variety of systems, including the dark matter (DM) halos of dwarf galaxies \citep[][]{Flores.Primack.94, Oman.etal.15, Agnello.Evans.12, Oh.etal.15}, low-surface-brightness galaxies \citep[][]{deBlok.etal.01, deBlok.Bosma.02, KuzioDeNaray.etal.08}, and massive star-forming disks at redshifts $z \sim 1-3$ \citep[][]{Genzel.etal.20, Price.etal.21}. 

The presence of DM cores has often been used as an argument against Cold Dark Matter (CDM), which predicts cuspy halos \citep[e.g.,][]{Navarro.etal.97}, and in favor of alternative dark matter models that predict halos to be cored, such as Self-Interacting Dark Matter \citep[SIDM, e.g.,][]{Spergel.Steinhardt.00, Kochanek.White.00, Rocha.etal.13, Tulin.Yu.18}. However, over the years, it has become clear that there are a variety of astrophysical processes that can transform a steep central cusp into a constant density core and/or remove dark matter from the central regions. Chief among these are explosive gas outflows driven by feedback from supernovae or active galactic nuclei (AGN), which cause dark matter and stars to expand in response to the fast shallowing of the central potential well. Supernova-driven outflows, especially if they are episodic \citep[][]{Pontzen.Governato.12, Teyssier.etal.13, ElZant.16, Li.etal.23}, are expected to be effective in low-mass galaxies \citep[][]{Dekel.Silk.86, DiCintio.etal.14, Tollet.etal.16, Freundlich.etal.20b}, while AGN-driven outflows may be relevant for creating cores in massive galaxies \citep[][]{Martizzi.etal.13, Dekel.etal.21}. Alternative methods for core creation are dynamical heating due to dynamical friction acting on a massive perturber \citep[][]{ElZant.etal.01, Goerdt.etal.10, Cole.etal.11, Ogiya.Nagai.22} or a bar \citep[][]{Weinberg.Katz.02}, tidal stripping of systems that are initially radially anisotropic \citep[][]{Chiang.vdBosch.25}, and core scouring due to a binary supermassive black hole \citep[][]{Begelman.etal.80, Quinlan.96}. The latter process is often invoked to explain the stellar cores observed in massive ellipticals \citep[][]{Ferrarese.etal.94, Byun.etal.96, Gebhardt.etal.96}. Hence, cores are a fairly natural by-product of structure formation, and their mere presence gives little to no insight regarding the nature of dark matter.

Numerical simulations of systems with central cores have revealed some intriguing dynamical phenomena. In particular, \cite{Read.etal.06c} noticed that a massive perturber that sinks due to dynamical friction toward the center of a system with a large constant density core stalls its inspiral close to the core radius. This `core stalling', which has since been reproduced in numerous  independent studies \citep[][]{Goerdt.etal.06, Inoue.09, Goerdt.etal.10, Inoue.11, Kaur.Sridhar.18, DuttaChowdhury.etal.19}, is not predicted by the standard \citet{Chandrasekhar.43} formulation of dynamical friction. Another dynamical phenomenon in cores, arguably even more confounding than core stalling, is `dynamical buoyancy', which was first noticed by \citet{Cole.etal.12}. Using $N$-body simulations to study what happens to a massive object that is initially located inside a constant density core, they found that the object starts moving outwards until it reaches the core radius, where its outward motion stalls. Finally, \citet{Dattathri.etal.25a} showed that cored systems can be unstable to the development of a rotating dipole mode. This mode saturates into a long-lived solitonic mode that can dislodge the entire central cusp, causing it to slosh back and forth through the central region of the galaxy without ever settling back down.

All these core-specific dynamical processes can have profound implications for various aspects of galaxy evolution. They can conspire to prevent massive objects such as nuclear star clusters and/or SMBHs from reaching, or remaining at, the center of their potential well, and may thus explain why in dwarf galaxies, largely believed to be cored, many nuclear star clusters and AGN are found to be offset from their center of mass \citep[][]{Bingelli.etal.00, Cote.etal.06, Shen.etal.19, Reines.etal.20, Sargent.etal.22, Mezcua.etal.24}. They can also prevent SMBHs from forming a hard binary, thereby inhibiting the SMBHs from merging together. Interestingly, mergers among SMBHs in dwarf galaxies are expected to dominate the gravitational wave signal to be measured with the Laser Interferometer Space Antenna (LISA) \citep[][]{Barause.etal.20, Volonteri.etal.20}. However, these estimates have not taken into account core stalling and might therefore be severely overestimated. Core stalling has also been invoked to explain the presence of multiple nuclei or globular clusters in the centers of galaxies, which would otherwise coalesce in a short time \citep[][]{Goerdt.etal.06, Bonfini.Graham.16, Boldrini.etal.19, SanchezSalcedo.Lora.22}. 

In a recent paper \citep[][hereafter D25]{Dattathri.etal.25b}, we presented a unified picture for core stalling, buoyancy, and the dipole instability operating in the cores of isotropic and spherically symmetric systems. Using idealized simulations and arguments based on kinetic theory, we showed that all these dynamical phenomena are related to specific features in the phase-space distribution function (DF) $f=f(E)$ of the host system. More specifically, core stalling occurs whenever the gradient $\rmd f/\rmd E$ at the orbital energy of the perturber vanishes (i.e., when the perturber reaches a plateau in the DF), while the dipole instability is triggered whenever the DF has an inflection, where $\rmd f/\rmd E > 0$. This mode exerts a torque on a central cusp or central object, such as a supermassive black hole or a nuclear star cluster, which can set it is motion, causing it to move outward, thus giving rise to dynamical buoyancy.

Bumps and plateaus in DFs may arise through various dynamical processes related to structure formation and evolution. In particular, \citetalias{Dattathri.etal.25b} show that inflections in the DF are common in spherical isotropic systems that have a rapid transition from a steep outer density profile to a shallow inner profile (see also \citealt{Weinberg.23}). Hence, processes that transform a central cusp in a core will affect the DF and might introduce local inflections that trigger the dipole instability. Moreover, dynamical friction acting on a massive perturber transfers energy and angular momentum from the perturber to the host system, thereby modifying its DF. Hence, even if the system {\it initially} has $\rmd f/\rmd E<0$ everywhere, the infall of a perturber can create a plateau or inflection.

So far, core stalling, buoyancy and the dipole instability have only been demonstrated in collisionless $N$-body systems, such as CDM halos, which obey the collisionless Boltmann equation (CBE). However, cores are a natural prediction of several alternative dark matter models. In particular, if dark matter is self-interacting, the resulting heat conduction gives rise to dark matter halos with large isothermal cores \citep[][]{Spergel.Steinhardt.00, Burkert.00}. Hence, it seems natural to postulate that massive perturbers sinking inside SIDM halos should stall near the core radius \citep[see for example][]{Adhikari.etal.25}. However, as shown in \citetalias{Dattathri.etal.25b}, not all cores are equal, in that the mere presence of a core does not guarantee core stalling and/or buoyancy. In this paper, we use high-resolution, idealized numerical experiments to examine how self-interactions impact core dynamics. Interestingly, we show that self-interactions suppress core stalling, buoyancy, and the dipole instability. If sufficiently strong, the self-interactions can completely suppress these phenomena such that, contrary to the case of CDM, massive objects rapidly sink to the center of the cored host system where they remain at rest. We also demonstrate that as the massive objects sink towards the center of an SIDM halo, they catalyze a drastically accelerated core collapse. These results offer a novel pathway to probing the nature of dark matter. 

This paper is organized as follows. In Section~\ref{sec:kinetics} we first give a brief overview of the dynamics underlying core stalling, buoyancy, and the dipole instability. Section~\ref{sec:method} presents the initial conditions for the different dark matter halos used throughout this study, the $N$-body code that we use, and its validation. Section~\ref{sec:stalling} studies core stalling in two different halos, comparing the results for CDM to those of SIDM. Dynamical buoyancy and the dipole instability are the topics of Section~\ref{sec:dipole}, again comparing how these processes manifest in CDM {\it vs.} SIDM. Section~\ref{sec:discussion} presents a discussion of why core dynamics in SIDM differ drastically from that in CDM. We end in Section~\ref{sec:concl} with a summary and discussion of the implications of our findings.

\section{Kinetic Theory of Core Dynamics}
\label{sec:kinetics}

Before assessing how core stalling, buoyancy, and the dipole instability are impacted by self-interactions, we first briefly review the unified kinetic picture for core dynamics advanced in \citetalias{Dattathri.etal.25b}. The key principle underlying all these core-specific dynamical processes is the secular transport of energy and angular momentum between the host particles and a perturbation. This perturbation might be a massive black hole that is sinking in due to dynamical friction, or it can be a point mode (also known as Landau mode) that is inherent to the system. The dipole instability is an example of the latter case\footnote{As shown in \citet{Weinberg.23} and \citet{Dattathri.etal.25a}, the dipole mode is a discrete point mode, also known as a Landau mode, which is a zero of the system's dispersion relation inferred from perturbing the coupled CBE-Poisson equation.}. According to kinetic theory \citep[see][for some excellent reviews]{Binney.Tremaine.08, Fouvry.17, Hamilton.Fouvry.24}, this secular transport occurs at resonances between the frequencies of the host particles and that of the perturber (i.e., the circular frequency of the inspiraling black hole or the eigen-frequency of the dipole mode). In particular, while particles on one side of the perturber gain energy from interacting with the perturber, those on the other side lose energy by transferring it to the perturber. Hence, whether the perturber or perturbation gains or loses energy depends on the net balance between these gainers and losers, which in turn is reflected by the gradient in the distribution function across the resonance. 

\subsection{Dynamical Friction and the LBK Torque}
\label{sec:LBKtorque}

Dynamical friction arises from a response induced in the DF of the host halo by the motion of a massive perturber. This response exerts a torque on the perturber that causes a transfer of angular momentum (and energy) between the perturber and the particles that make up the host halo. Applying linear perturbation theory to the CBE-Poisson system that describes a collisionless gravitational system, \citet{Tremaine.Weinberg.84} showed that (in the time-asymptotic limit, $t \rightarrow \infty$), this torque is given by 
\begin{equation}
\label{LBK1}
\calT_{\rm LBK} = 16 \pi^4 \sum_{\boldsymbol{\ell}} \ell_3 \int \rmd \vect{J} \, \delta(\boldsymbol{\ell} \cdot \vect{\Omega} - \ell_3 \Omega_\rmp) \, \boldsymbol{\ell} \cdot \frac{\partial f}{\partial \vect{J}} \, \left| \hat{\Phi}_{\boldsymbol{\ell}}(\vect{J}) \right|^2\,.
\end{equation}
which is known as the LBK torque after \citet{LyndenBell.Kalnajs.72} who first derived it in their treatment of angular momentum transport due to spiral structure in disk galaxies. Here, $\vect{J}=(J_1,J_2,J_3)$ are the three actions, $\vect{\Omega} = (\Omega_1,\Omega_2,\Omega_3)$ are the corresponding frequencies, $\boldsymbol{\ell}=(\ell_1,\ell_2,\ell_3)$ is an integer vector, $\Omega_\rmp$ is the frequency of the perturber, $\left| \hat{\Phi}_l(\vect{J}) \right|$ are the Fourier modes of the perturber potential, $\delta(x)$ is the Dirac delta function, and $f=f(\vect{J})$ is the unperturbed DF of the halo. The summation runs from $-\infty$ to $\infty$ for $\ell_1$ and $\ell_2$, and from $0$ to $\infty$ for $\ell_3$. In the case of a spherical isotropic system, the DF is a one-dimensional function of energy, so the LBK torque reduces to
\begin{equation}
\label{LBK2}
\calT_{\rm LBK} = 16 \pi^4 \Omega_\rmp \sum_{\boldsymbol{\ell}} \ell^2_3 \int \rmd \vect{J} \, \delta(\boldsymbol{\ell} \cdot \vect{\Omega} - \ell_3 \Omega_\rmp) \, \frac{\rmd f}{\rmd E} \, \left| \hat{\Phi}_{\boldsymbol{\ell}}(\vect{J}) \right|^2\,.
\end{equation}
This expression clearly exposes that the torque on the perturber arises from orbits that are on resonance with the perturber (i.e., that satisfy the commensurability condition $\boldsymbol{\ell} \cdot \vect{\Omega} - \ell_3 \Omega_\rmp = 0$) and that it is the gradient of the DF in action space at those resonances that determines the sign of the torque and thus the net direction of flux transfer.

Spherical isotropic systems that have $\rmd f/\rmd E<0$ at all energies are stable to both radial and non-radial modes \citep[][]{Antonov.62, Doremus.etal.73}. As is apparent from equation~(\ref{LBK2}), the LBK torque in such systems is always negative (that is, retarding) and the perturber thus experiences regular dynamical friction. What makes kinetic theory of gravitational systems complicated is that there are typically an infinite number of resonances (represented by the summations in equations~[\ref{LBK1}]-[\ref{LBK2}]), and at each resonance there are typically three gradients of the DF with respect to three actions. However, \citet{Kaur.Sridhar.18} have shown that within the core region of a spherical isotropic system, the LBK torque is dominated by the contribution from the corotation resonance (CR), where the azimuthal frequency of the particles matches that of the perturber \citep[see also][]{Kaur.Stone.22, Banik.vdBosch.22}. Therefore, the sign of the net torque is determined, to a good approximation, by the gradient $\rmd f/\rmd E$ at the energy associated with co-rotation, which we refer to as $(\nabla f)_{\rm CR}$ in what follows. If $(\nabla f)_{\rm CR} < 0$, then the LBK torque is retarding and the perturber experiences regular dynamical friction. However, if $(\nabla f)_{\rm CR} > 0$, the torque is enhancing, causing the perturber to gain energy and move outward, which is a manifestation of dynamical buoyancy. Finally, if $(\nabla f)_{\rm CR} = 0$ the torque vanishes and the perturber is expected to undergo core stalling. Hence, the expression for the LBK torque exemplifies that core stalling and buoyancy have a natural explanation in kinetic theory and should be linked to the detailed shape of the DF of the host system. Using a series of $N$-body experiments, \citetalias{Dattathri.etal.25b} established that dynamical friction, core stalling and buoyancy can indeed be linked to the sign of $(\nabla f)_{\rm CR}$, at least for a massive perturber in a spherical background halo with an (initially) isotropic DF. 

Although this simple picture based on the LBK torque offers a natural explanation for core stalling that is consistent with the results of numerical simulations, it only captures part of the full dynamics. Here, we summarize some of the complications and refer the reader to \citetalias{Dattathri.etal.25b} for a more detailed discussion. Firstly, the LBK torque is based on linear perturbation theory and therefore does not account for non-linear effects such as orbit trapping. Therefore, it is only valid in what \citet{Tremaine.Weinberg.84} refer to as the ``fast regime'', when the perturber is sinking fast enough such that it sweeps through the resonances without any of them building up to non-linear amplitudes. Hence, the LBK torque is only of limited use in describing core-stalling, which transpires in the ``slow regime'', in which particles can get trapped onto librating orbits that continuously and periodically exchange energy and angular momentum with the perturber \citep{Tremaine.Weinberg.84, Chiba.Schonrich.22, Chiba.23, Banik.vdBosch.22}. Over time, phase-mixing among these librating orbits will rapidly drive their net action flux to zero. Hence, if the libration zone is evolving sufficiently slowly over time, the librating orbits contribute a negligible torque and the perturber will stall. Secondly, the LBK torque is calculated based on the {\it unperturbed} DF and does not account for the evolution in the DF due to secular flux transfer as the perturber sinks in. Quasi-linear theory \citep[e.g.,][]{Weinberg.01, Chavanis.24} dictates that secular flux transfer causes a diffusion in phase space that, given enough time, generates a plateau in the DF $(\rmd f/\rmd E = 0)$ at the location of each resonance \citep[][]{Hamilton.24, Banik.Bhattacharjee.25}. In cored systems, the co-rotation resonance dominates; therefore, a slowly sinking perturber can carve out a strong plateau as it enters the core, suppressing the LBK torque and halting further inspiral. Hence, dynamical friction can be self-limiting in that the sinking in of the perturber can modify the DF to the point that it induces stalling. Indeed, such behavior has been observed in $N$-body simulations \citep[][see also Section~\ref{sec:DFstall} below]{Goerdt.etal.10, Dattathri.etal.25b}. Finally, the LBK torque is only valid in the asymptotic limit ($t \rightarrow \infty$) of adiabatic growth of the perturber. However, if the system happens to be unstable, then the developing instabilities are likely to become important before the LBK torque has a change to develop. As discussed in Section~\ref{sec:dipole}, this is likely the case with dynamical buoyancy. Despite these details, \citetalias{Dattathri.etal.25b} show that both in the linear and weakly non-linear regimes, the gradient in the DF is indeed the key determinant in the magnitude and direction of frictional force, and we adhere to this framework in this paper.

\subsection{Dipole Instability and Dynamical Buoyancy}
\label{sec:dipmode}

Gravitational $N$-body systems have a spectrum of modes at discrete frequencies (referred to as point modes or Landau modes), which describe perturbations to the system that result in a response that has the same functional form as the perturbation. These modes, of which the dipole mode is a typical example, are continuously excited, either by external perturbations or by discreteness noise inherent to the system. Whether they grow or decay depends on the gradients in the DF at the resonances between the eigen-frequency of the mode and the orbital frequencies of the particles of the system in question. 

Although the dipole mode is common, it is typically (weakly) damped \citep[][]{Weinberg.94c, Heggie.etal.20}. The first to report an example of a system with an unstable (growing) dipole mode was \citet{Weinberg.23}. He noticed the instability in a spherical isotropic NFW halo, commonly used to model cold dark matter halos, but only if he truncated the density profile at large radii. This truncation induces an inflection in the DF (i.e., a region in energy where $\rmd f/\rmd E>0$), thus violating Antonov's stability criterion. Recently, \citet[][]{Dattathri.etal.25a} showed that spherical isotropic systems with a double power-law density profile that have a rapid transition from a steep outer density gradient to a shallow inner density gradient have a similar inflection in their DFs. Similarly to the system studied by \citet{Weinberg.23}, these systems are unstable to the development of a rotating dipole mode, whose growth is facilitated by the positive gradient in the DF. 

As the dipole instability grows in amplitude, it ultimately saturates developing into a long-lived solitonic mode, which is the dipolar equivalent of a quadrupolar bar mode in a disk galaxy. Most relevant for this paper, \citet{Dattathri.etal.25a} showed that this mode exerts a torque that can dislodge a central cusp, setting it in motion about the system center. Over time, a long-lived dynamical equilibrium is established in which the cusp continues to slosh back and forth through the central region of the galaxy. In \citetalias{Dattathri.etal.25b}, we have shown that the same mechanism can also dislodge a massive perturber from the center. Hence, dynamical buoyancy is a manifestation of the dipole instability which, in turn, has its origin in an inflection in the DF.

\section{Methodology}
\label{sec:method}

The main goal of this paper is to investigate whether core stalling and/or dynamical buoyancy occur in the cores of SIDM halos. Using idealized $N$-body simulations of isolated dark matter halos in which we introduce a massive perturber, we study how the perturber spirals inwards due to dynamical friction, and how the time scale thereof depends on the collisional cross section of the dark matter particles. Crucially, we contrast the findings with those in standard CDM (i.e., zero cross-section for self-interactions).
\begin{figure*}
\centering
\includegraphics[width=\textwidth]{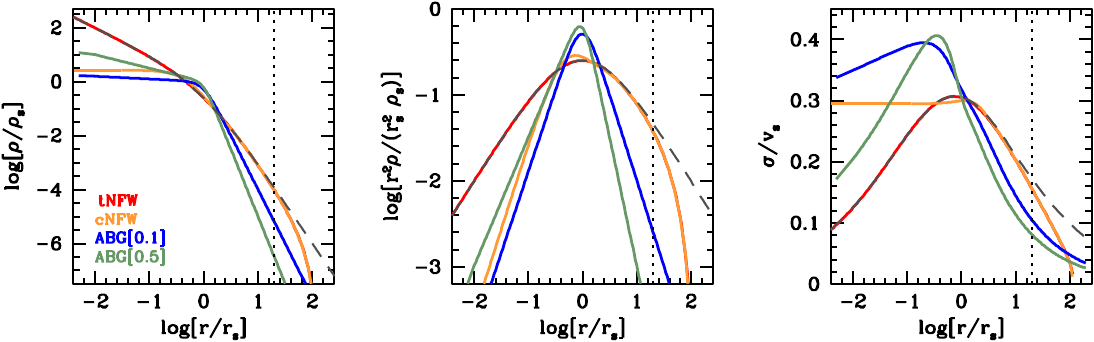}
\caption{Initial density and velocity dispersion profiles for the various halos discussed in this paper, as indicated. The gray dashed profiles correspond to the untruncated NFW profile and is shown for comparison. Note that the density profile of the truncated NFW is almost indistinguishable from the untruncated NFW profile inside the virial radius, indicated as the vertical, dotted line.}
\label{fig:initprofs}
\end{figure*}

\subsection{Host halos}
\label{sec:hosts}

In this paper, we consider three different host halos. All are assumed to be spherical and to have an isotropic distribution function (DF) $f=f(E)$, but they have different density profiles. The first is a cored, truncated NFW \citep{Navarro.etal.97} halo, in which the core is created due to self-interactions. This therefore represents a typical halo one might expect if dark matter is indeed self-interacting. The second halo has a double-power law density profile with an inner core that has a shallow density gradient ($\rho \propto r^{-0.1}$). This halo is chosen because it is known, in the case of CDM, to exhibit strong core stalling at a relatively large radius. Hence, this halo is an ideal benchmark for testing how core stalling depends on the self-interaction cross section. The third and final halo has a similar double-power law density profile as halo two, but with a central $r^{-0.5}$ cusp. We use this halo because \citet{Dattathri.etal.25a} has shown that it is highly unstable to the development of a dipole instability, at least in the case of CDM. Hence, we will use this halo to study how the dipole instability and the associated dynamical buoyancy behave in the presence of self-interactions. In what follows, we refer to these three halos as the cNFW, \Core and \Cusp halos, respectively. Here `ABG' refers to the parameters $\alpha$, $\beta$ and $\gamma$ that characterize the double power-law density profile (see below)

The initial conditions for the cNFW system are created as follows. We first initialize a halo with an NFW density profile,
\begin{equation}
\rho(r) = {\rho_\rms \over (r/r_\rms) (1+r/r_\rms)^2}\,,
\end{equation}
with scale radius $r_\rms$ and characteristic density $\rho_\rms$, which we truncate using the energy-truncation method of \citet{Drakos.etal.17}\footnote{This is required since an untruncated NFW profile has infinite mass, and since a hard-truncation at the virial radius, or any other radius, results in instability \citep[][]{Weinberg.23}.}. This is similar to the method used to create King models by modifying the DF of an isothermal sphere \citep[][]{Michie.63, King.66, Binney.Tremaine.08}. Specifically, we use an energy value for the truncation such that the total truncated mass $M_{\rm tot} = 1.2703 \, M_{\rm vir}$. This is chosen such that the density profile inside the virial radius, $r_{\rm vir}$, is almost identical to that of the untruncated profile. Here $M_{\rm vir}$ is the virial mass of the system, defined as the mass enclosed by the {\it untruncated} NFW halo inside $r_{\rm vir}$, defined as the radius inside which the average density is $\Delta_{\rm vir}$ times the critical density for closure. Throughout, we adopt $\Delta_{\rm vir}=97$, in accordance with the Planck cosmology \citep[][]{Bryan.Norman.98, Planck20}, and a halo concentration parameter $c=\rvir/r_\rms=20$. The red curves in Fig.~\ref{fig:initprofs} show the density and 1D velocity dispersion profiles of this truncated NFW (hereafter tNFW) profile, normalized to $\rho_\rms$ and $v_\rms \equiv \sqrt{G M_\rms/r_\rms}$, respectively, where the characteristic mass $M_\rms \equiv 4 \pi \rho_\rms \, r_\rms^3$. For comparison, the gray-dashed lines show the corresponding untruncated profiles. Note that the tNFW system looks very similar to the untruncated NFW profile out to the virial radius, with $M(<\Rvir) \sim 0.95 \Mvir$. 

As described in detail in Section~\ref{sec:validation}, the tNFW halo is subsequently evolved in isolation assuming self-interacting dark matter with a cross section per unit mass equal to $\sigma_\rmm = 25 \cmg$, until it reaches the epoch of maximum core (i.e., minimal central density, shortly before the onset of core collapse). This cored-out version of the truncated NFW halo is representative of a typical evolved halo if dark matter has non-negligible self-interactions and serves as the initial conditions for our subsequent numerical experiments described below. The orange curves in Fig.~\ref{fig:initprofs} show the corresponding density and velocity dispersion profiles of this cored-out NFW profile, to which we refer as cNFW hereafter. Note how the cNFW system is indistinguishable from the tNFW halo for $r \gta r_\rms$. Inside of the scale-radius, however, the self-interactions have created a pronounced isothermal core of constant density. 

The \Core and \Cusp halos have a double power-law density profile characterized by the general $\abg$-profile \citep{Zhao.96},
\begin{equation}
\label{abgprof}
    \rho(r) = \rho_\rms \, \left( \frac{r}{r_\rms} \right)^{-\gamma} \left[ 1 + \left( \frac{r}{r_\rms} \right)^{\alpha} \right]^{\frac{\gamma-\beta}{\alpha}}  \,.
\end{equation}
This profile transitions from an outer power-law, $\rho \propto r^{-\beta}$, to an inner power-law, $\rho \propto r^{-\gamma}$, with $\alpha$ controlling the steepness of the transition. Several commonly encountered density profiles in astrophysics are examples of this $\abg$ profile, including the (untruncated) NFW profile, which has $\abg=(1,3,1)$, and the \citet{Hernquist.90}, \citet{Jaffe.83} and \citet{Plummer.11} profiles, which have $\abg=(1,4,1)$, $(1,4,2)$ and $(2,5,0)$, respectively. For the \Core halo we adopt $\abg = (4,4,0.1)$, which transitions from an outer power-law with $\rho \propto r^{-4}$ to an inner core-like region with $\rho \propto r^{-0.1}$. The corresponding density and 1D velocity dispersion profile, normalized by $\rho_\rms$ and $v_\rms$, respectively, are shown as the blue curves in Fig.~\ref{fig:initprofs}. Note that the core region of this halo is not isothermal, unlike the cNFW halo. As shown in \citetalias{Dattathri.etal.25b} a CDM halo with this density profile causes massive perturbers to experience core stalling at a relatively large radius (see also Section~\ref{sec:DFcore} below). 

Finally, the density profile of the \Cusp halo also follows the $\abg$-profile of equation~(\ref{abgprof}), but with $\abg = (6,5,0.5)$. Hence, its density profile rapidly transitions from $\rho \propto r^{-5}$ in the outskirts to a central cusp with $\rho \propto r^{-0.5}$. Its density and velocity dispersion profile are shown as the green curves in Fig.~\ref{fig:initprofs}. \citet{Dattathri.etal.25a} have shown that a CDM halo with this density profile is unstable to a dipole mode that dislodges the central cusp from the center.

Throughout, we adopt model units for which the gravitational constant, $G$, the total mass of the system, $M_{\rm tot}$, and the scale radius, $r_\rms$, are all unity. For the tNFW halo, we instead set $G=\Mvir=r_\rms=1$, which implies a total mass of $M_{\rm tot} = 1.2703$. When converting to physical units, we adopt the following fiducial scales: For the tNFW halo, we set $M_{\rm vir} = 3\times 10^9 \msunh$, and we adopt a concentration parameter $c=20$. Throughout we adopt a Hubble parameter $h=H_0/100\kmsmpc = 0.7$, which implies that $r_\rms = 2.13\kpc$. For the \Core and \Cusp halos, we set $M_{\rm tot} = 3\times 10^9 \msunh$ while their characteristic densities $\rho_\rms$ are set to be equal to that of the tNFW halo ($1.69 \times 10^{-2} \Msun/\pc^3$). This implies that their scale radii, $r_\rms$, are $1.23$ and $1.39$  times larger, respectively, than that of the tNFW halo. This ensures that all three halos have the same characteristic core-crossing time $t_{\rm core} \equiv r_\rms/v_\rms = 32 \Myr$. 

In the case of SIDM, where relaxation is driven by short-range interactions resulting in large-angle deflections, another relevant time scale is the characteristic collision time between individual collisions, which is given by
\begin{equation}\label{tcoll}
 t_0 = (a \rho_\rms \, \sigma_\rmm \, v_\rms)^{-1}\,,
\end{equation}
\citep[][]{Balberg.etal.02}, where $a = 4/\sqrt{\pi} \approx 2.26$ for elastic hard-sphere-like interactions as assumed here. For a cross section of $\sigma_\rmm = 1 \cmg$, our fiducial halos have characteristic collision times of $t_0 = 1.91 \Gyr$ (tNFW), $1.55 \Gyr$ (\Core) and $1.37 \Gyr$ (\Cusp), respectively, which are fairly comparable to the halo crossing times, $t_{\rm cross} = \rvir/\Vvir$, where $\Vvir = \sqrt{G \Mvir/\rvir}$. Hence, for a cross section of $1 \cmg$, a particle has of order one collision per crossing time. 
\begin{figure*}
\centering
\includegraphics[width=\textwidth]{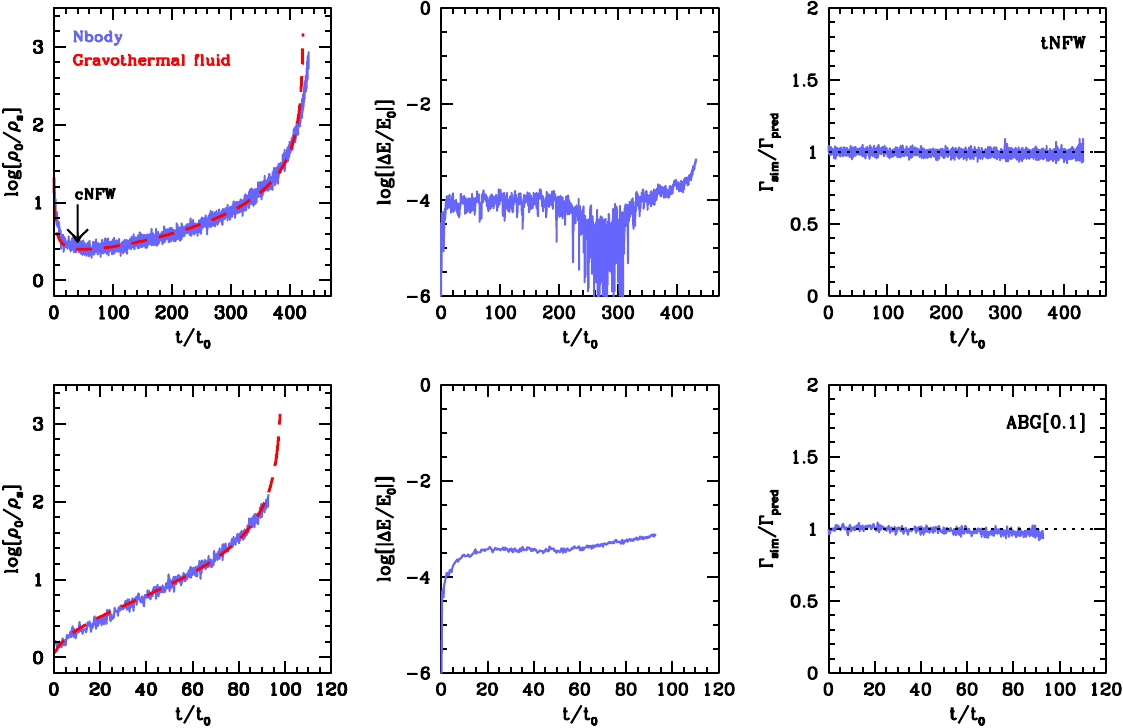}
  \caption{Evolution of two SIDM halos ($\sigma_\rmm = 25 \cmg$) as a function of time, expressed in units of the collision time $t_0$. From left to right the panels show the evolution of the central density, the change in the total energy of the system normalized by the initial energy, and the ratio between the scattering rate in the $N$-body simulation, $\Gamma_{\rm sim}$, and the predicted rate, $\Gamma_{\rm pred}$, based on the instantaneous density and velocity dispersion (equation~[\ref{GammaExp}]). Top and bottom panels are for the tNFW and \Core halos, respectively. The red dashed lines in the left-hand panels are the predictions based on the gravothermal fluid equations (see Appendix~\ref{sec:gravothermal}). The downward arrow in the top-left panel marks the time of maximum core, which is used as the initial conditions for the cNFW simulations.} 
\label{fig:tNFWevol}
\end{figure*}

Finally, because of the self-similar nature of gravothermal core collapse \citep[][]{Balberg.etal.02}, one may rescale our simulations to other combinations of total halo mass and collisional cross section following the guidelines presented at the end of Section~\ref{sec:sims}.

\subsection{Initial Conditions}
\label{sec:ICs}

Each host halo is sampled using $N=10^6$ particles, for which we draw phase-space coordinates using the corresponding isotropic DF $f = f(E)$, which is computed using the standard Eddington inversion method \citep[][]{Eddington.16, Binney.Tremaine.08}.

Physical systems must have a positive phase-space density, i.e. $f(E)>0$ at all $E$. In addition, according to Antonov's stability theorem, the system is stable if $\rmd f/\rmd E <0$ at all $E$ \citep{Antonov.62, Henon.73, Binney.Tremaine.08}. These conditions are satisfied for the tNFW and cNFW halos, but not for the \Core and \Cusp halos. In fact, as shown by \citet{Dattathri.etal.25a} both have an inflection in their DF where $\rmd f/\rmd E>0$ (see also Section~\ref{sec:discussion}). Using both linear modal analysis and numerical simulations (assuming collisionless particles), \citet{Dattathri.etal.25a} showed that such systems are unstable to a dipole mode in the center of the halo. This dipole mode saturates, becoming a long-lived solitonic mode, which is the dipolar equivalent of a quadrupolar bar mode in a disk galaxy. As part of our study, we investigate whether the same instability is present in the presence of self-interactions (see Section~\ref{sec:dipole} below).
\begin{figure*}
\centering
\includegraphics[width=0.95\textwidth]{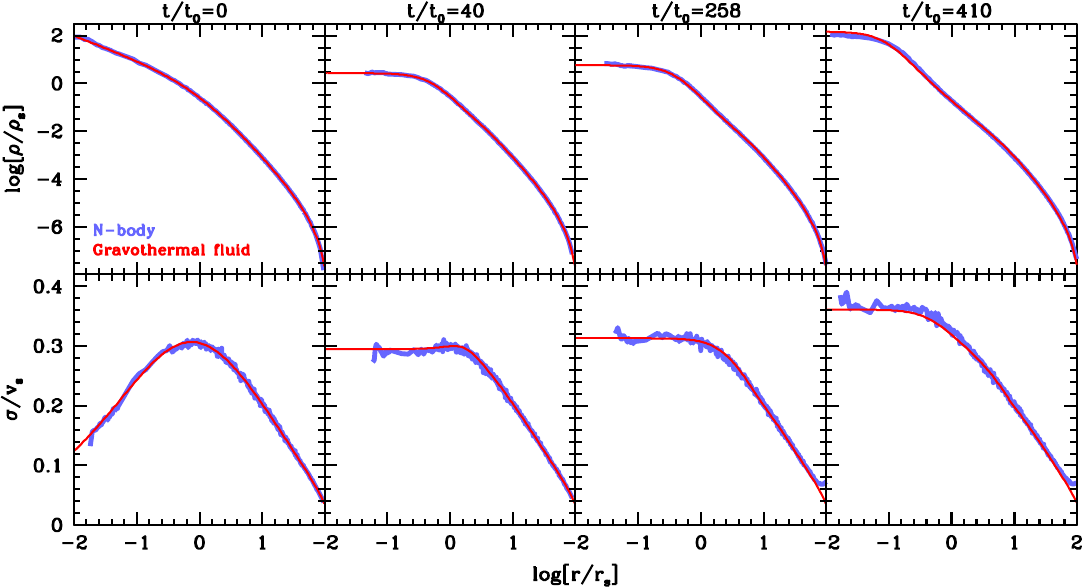}
  \caption{Evolution of the tNFW halo using SIDM with $\sigma_\rmm = 25\cmg$. Top and bottom panels show the density profile and the 1D velocity dispersion profile, respectively, both normalized to their characteristic values. Different columns correspond to different times, as indicated at the top. In each panel, blue curves show the results from the $N$-body simulation, while the red lines show the predictions based on the gravothermal fluid code. Note the excellent agreement between the two even for $t/t_0=410$, which is well into the core collapse regime (cf. Fig.~\ref{fig:tNFWevol}). The results at $t/t_0=40$, shown in the second column, serve as the initial conditions for the cNFW halo used in Section~\ref{sec:DFtNFW}.} 
\label{fig:tNFWprofs}
\end{figure*}

\subsection{$N$-Body Simulations}
\label{sec:sims}

All simulations are run using a modified version of the hierarchical $N$-body code {\tt treecode}, written by Joshua Barnes. {\tt treecode} uses a \cite{Barnes.Hut.86} octree to compute accelerations based on a multipole expansion up to quadrupole order, and a straightforward second order leap-frog integration scheme to solve the equations of motion. The forces between the particles are softened using a simple Plummer softening. 

To account for DM self-interactions, we closely follow \citet{Kochanek.White.00} and \citet{Vogelsberger.etal.12}. We assume that scattering is isotropic and elastic, and characterized by a velocity-independent cross section per unit mass\footnote{for more elaborate models for self-interactions, see e.g. \citet{Tulin.etal.13}, \citet{Robertson.etal.17}, and \citet{Arido.etal.25}.} equal to $\sigma_\rmm$, and that a particle $i$ can only scatter with one of its $k=32$ nearest neighbours. We compute the expectation value for the number of scattering events of particle $i$ in a time step $\Delta t$ according to
\begin{equation}\label{Nscat_expected}
\langle N_{{\rm scat},i} \rangle = \frac{1}{2} \sum_j \sigma_m \, m_\rmp \, W(r_{ij}, h_i) \, |\vect{v}_j - \vect{v}_i| \, \Delta t\,,
\end{equation}
where the summation is over the $k$ nearest neighbours, $m_\rmp$ is the particle mass, $W(r,h)$ is a kernel function that describes the `extent' of the simulation particle in configuration space, and $h_i$ is the smoothing length enclosing the $k$ nearest neighbours of particle $i$. The factor $1/2$ is needed to avoid double counting of the scattering events. The actual number of interactions for particle $i$ is drawn from a Poisson distribution with an expectation value equal to $\langle N_{{\rm scat},i} \rangle$. Ideally, $\Delta t$ is adjusted such that the probability for $N_{\rm scat}>1$ is sufficiently small, which requires that 
\begin{equation}\label{timestepSI}
 \Delta t < (\Delta t)_{\rm SI} \equiv \frac{\kappa}{{\rm max} (\frac{1}{2} \, \rho_i \, \sigma_\rmm \, \langle v_{\rm rel} \rangle_i)}   
\end{equation}
Here $\kappa$ is a free parameter, $\rho_i = m_\rmp \sum_j W(r_{ij},h_i)$, $\langle v_{\rm rel} \rangle_i = {1 \over k} \sum_j |\vect{v}_j - \vect{v}_i|$ where the sum is over the $k$ nearest neighbours of particle $i$, and the maximum in the denominator is taken over all $N$ simulation particles $i$. It is common to pick a value for $\kappa$ such that the probability of a particle scattering more than once per time step is negligible. However, smaller values for $\kappa$ make the simulation run slower, especially if, as in our case, the same time step is adopted for all particles.  We have experimented extensively with different values of $\kappa$ and find that even adopting values as large as $\kappa=0.5$ yields results that are indistinguishable. In fact, in a forthcoming paper (van den Bosch et al. 2026, in prep.) we explicitly demonstrate that having multiple scattering events per time step is not a major concern as long as one properly interprets equation~(\ref{Nscat_expected}) as the expectation value of a Poisson distribution, rather than as a probability \citep[see][for a discussion]{Fischer.etal.24}.

Once it has been determined that a particle $i$ has $N_{\rm scat}\geq 1$ scattering events in time step $\Delta t$, we proceed as follows. For each scattering event, particle $i$ is scattered against one of its 32 nearest neighbours, where the probability to scatter with particle $j$ is given by
\begin{equation}\label{Pscatter}
P_{ij} = \frac{W(r_{ij}, h_i) \, |\vect{v}_j - \vect{v}_i|}{\sum_j W(r_{ij}, h_i) \, |\vect{v}_j - \vect{v}_i|}\,,
\end{equation}
Throughout, we assume elastic scattering. Hence, once a particle $j$ is selected for a collision with $i$, we update their velocities according to\footnote{This is only valid if all particles have equal mass, which is the case in all our experiments.}
\begin{equation}\label{updatevel}
\vect{v}_i = \vect{v}_{\rm cm} + (\vect{v}_{ij}/2) \, \hat{\vect{e}} \, \nonumber 
\end{equation}
\vskip -0.55truecm
\begin{equation}
\vect{v}_j = \vect{v}_{\rm cm} - (\vect{v}_{ij}/2) \, \hat{\vect{e}} \,,
\end{equation}
where $\vect{v}_{ij} = \vect{v}_j - \vect{v}_i$, $\vect{v}_{\rm cm}$ is the center-of-mass velocity of the pair, and $\hat{\vect{e}}$ is a unit directional vector that we draw randomly from the unit sphere. Note that this procedure conserves both energy and linear momentum. If $N_{\rm scat}$ is larger than unity, we repeat this procedure without updating the probabilities in equation~(\ref{Pscatter}) but using the updated velocities when computing $\vect{v}_{ij}$ and $\vect{v}_{\rm cm}$ in equation~(\ref{updatevel}). If the selected target particle $j$ has already scattered with $i$ during the time step in question, we redraw a new target from $P_{ij}$ such that multiple scattering between identical pairs within the same time step is avoided. We have verified, though, that not redrawing yields results that are indistinguishable.

For our fiducial method, we adopt the standard (normalized) cubic spline function
\begin{equation}\label{spline}
W(r,h) = \frac{8}{\pi h^3} \left\{ \begin{array}{ll}
1 - 6q^2 - 6 q^3 & 0 \leq q \leq \frac{1}{2} \\
2(1 - q)^3       & \frac{1}{2} < q \leq 1 \\
0                & q > 1
\end{array}
\right. \,,
\end{equation}
where $q=r/h$. However, we have also experimented with a top-hat kernel, setting 
\begin{equation}\label{tophat}
W(r,h) = \frac{3}{4 \pi h^3} \left\{ \begin{array}{ll}
1 & q \leq 1 \\
0 & q > 1
\end{array}
\right. \,,
\end{equation}
which is basically the approach used by \citet{Kochanek.White.00}, and find results that are indistinguishable \citep[see also][]{Fischer.etal.21}.

All particles advance with the same time step $\Delta t$, whose value is adjusted each time step so that
\begin{equation}\label{timestep}
\Delta t = {\rm min}\left[ (\Delta t)_{\rm SI}, (\Delta t)_{\rm grav} \right]\,.
\end{equation}
Here $(\Delta t)_{\rm SI}$ is the time step for the self-interactions given by equation~(\ref{timestepSI}) and 
\begin{equation}\label{timestepgrav}
 (\Delta t)_{\rm grav} = \sqrt{2 \, \eta \, \frac{\epsilon}{{\rm max} |\vect{a}_i|}}
\end{equation}
is the time step used for the gravity solver. Here $\vect{a}_i$ is the gravitational acceleration of particle $i$, $\epsilon$ is the Plummer softening length, and $\eta$ is a free parameter. Throughout, we use $\epsilon = 0.0232$ and $\eta = 0.025$, and we have verified that our results are stable to modest changes in these parameters. Although adjusting $\Delta t$ each time implies a loss of symplecticity of our integration scheme \citep[][]{Dehnen.17}, we find that the total energy in our simulations is conserved to better than one part in $10^3$ up to the final stages of core collapse (see Section~\ref{sec:validation} below) \footnote{Note that even for a perfectly symplectic integrator total energy is not exactly conserved \citep[][]{Wisdom.Holman.91}.}

A useful measure to quantify self-interactions is the (local) Knudsen number, $\Kn$, defined as the ratio of the mean free path, $\lambda_{\rm mfp} = (\rho \sigma_\rmm)^{-1}$, and the gravitational scale height (or Jeans length), $H = \sigma_v / \sqrt{4 \pi G \rho}$, where $\sigma_v$ is the local velocity dispersion. We have verified that all our simulations are at all time in the long mean-free-path (LMFP) regime, for which $\Kn \equiv \lambda_{\rm mfp}/H > 1$ (see Appendix~\ref{sec:Knudsen}). In this regime, the evolution of the system is self-similar when expressed as a function of $t/t_0$ \citep[][]{Balberg.etal.02, Koda.Shapiro.11, Yang.etal.23}. 

For all our simulations, the collisional cross section in model units (see Section~\ref{sec:hosts}) is given by
\begin{equation}\label{tildesigma}
\tilde{\sigma}_\rmm = 0.121 \, Q_{\alpha\beta\gamma}^{2/3} \, \left({\sigma_\rmm \over \cmg}\right) \, \left({M_\rmh \over 3 \times 10^{9} \Msunh}\right)^{1/3} \,,
\end{equation}
where,
\begin{equation}
Q_{\alpha\beta\gamma} = \int_0^{x_{\rm max}} x^{2-\gamma} \, \left(1 + x^{\alpha}\right)^{(\gamma-\beta)/\alpha} \, \rmd x\,,
\end{equation}
with $\alpha$, $\beta$ and $\gamma$ defined in equation~(4) of the main text. The upper integration bound $x_{\rm max} = c = 20$ in the case of the cNFW halo and $x_{\rm max} = \infty$ for both the \Core and \Cusp  halos. The halo mass $M_\rmh$ in equation~(\ref{tildesigma}) is equal to $M_{\rm vir}$ in the case of the cNFW simulations and to $M_{\rm tot}$ for the two ABG halos.  Since the Knudsen number is inversely proportional to $\sigma_\rmm M_\rmh^{1/3}$, and thus to $\tilde{\sigma}_\rmm$, one may rescale our simulations to any other combination of $\sigma_\rmm$ and $M_\rmh$ that leaves $\tilde{\sigma}_\rmm$ invariant.

\subsection{Validation of self-interactions}
\label{sec:validation}

In order to ensure that our implementation of the self-interactions is adequate, we evolve the systems in isolation (i.e., no perturber) and (i) compare the scatter rate in the simulation to theoretical predictions and (ii) compare the evolution of the density and velocity dispersion profiles with predictions computed using a gravothermal fluid model.

The scattering rate of a halo is predicted to be
\begin{equation}\label{GammaExp}
\begin{split}
\Gamma_{\rm pred} &= \frac{1}{2} \int n(\vect{x}) \, \rho(\vect{x}) \, \sigma_\rmm \, \langle v_{\rm rel} \rangle(\vect{x}) \, \rmd^3\vect{x} \\
&= 8 \sqrt{\pi} \, N_\rmp \, \sigma_\rmm \, \frac{1}{M_{\rm tot}} \int_0^{\infty} \rho^2(r) \, \sigma_{\rm 1D}(r) \, r^2 \, \rmd r\,,
\end{split}
\end{equation}
where $n(\vect{x}) = \rho(\vect{x})/m_\rmp$ is the number density of particles, $\sigma_{\rm 1D}(\vect{x})$ is the local one-dimensional velocity dispersion, and in the second step we have assumed spherical symmetry and that the particles follow a Maxwell-Boltzmann (MB) distribution. Although we are aware that the velocities of dark matter particles are not expected to follow a MB distribution, this oversimplification is not expected to have a large effect. We evaluate $\Gamma_{\rm pred}$ every few time steps using the instantaneous density and velocity dispersion profiles of the halo computed on the fly in the simulation code, and compare its value against the actual scattering rate in the simulation. The latter is evaluated as $\Gamma_{\rm sim} = (\sum_i N_{{\rm scat},i})/\Delta t$, where $N_{{\rm scat},i}$ is the number of scattering events that particle $i$ experienced during the time period $\Delta t$. As we demonstrate below, our simulations consistently have $\Gamma_{\rm sim}/\Gamma_{\rm pred} = 1$ within the shot noise.

As shown by numerous authors, the evolution of a spherical SIDM halo can be described using the conducting gravothermal fluid model first developed by \citet{LyndenBell.Eggleton.80} to describe the collisional evolution of globular clusters \citep[][]{Balberg.etal.02, Koda.Shapiro.11, Pollack.etal.15, Essig.etal.19, Nishikawa.etal.20}. Appendix~\ref{sec:gravothermal} describes our method of solving these gravothermal fluid equations, which closely follows the treatment of \citet{Nishikawa.etal.20}. Fig.~\ref{fig:tNFWevol} shows results for SIDM simulations with $\sigma_\rmm = 25\cmg$ for both the tNFW halo (upper panels) and the \Core halo (lower panels). The left-hand panels show the evolution of the central density as a function of time, $t$, normalized by the characteristic collision time, $t_0$, given by equation~(\ref{tcoll}). For comparison, the red dashed lines show the predictions from the gravothermal fluid model, which are in excellent agreement with the simulation results. We caution that this agreement is only achieved after tuning one free parameter that characterizes the conductivity in the LMFP regime (see equation~[\ref{condLMFP}]). Note that the tNFW and \Core halo evolve rather differently. Initially, the central density of the tNFW halo declines rapidly, reflecting the phase of core formation during which the $r^{-1}$ cusp of the initial NFW profile is eroded away due to the self-interactions that cause a net flux of energy from the halo outskirts towards the center. The direction of this flux is dictated by the radial gradient of the velocity dispersion, which, for an NFW halo, is positive for $r \lta r_\rms$ (see right-hand panel of Fig.~\ref{fig:initprofs}). Once the central gradient in the velocity dispersion has been removed (at $t/t_0 \sim 30$), and an isothermal core has been established, the energy flux due to self-interactions reverses sign, as now the central region is dynamically hotter than the halo outskirts. This triggers core collapse through the gravothermal catastrophe, resulting in an increase in central core density at an ever-increasing rate \citep[e.g.,][]{LyndenBell.Wood.68, Balberg.etal.02, Koda.Shapiro.11}. During core collapse, both the central density and the central velocity dispersion increase with time, resulting in a drastic reduction of the simulation time steps causing the simulation to crawl to a halt. Therefore, we typically terminate the simulation once the central core density exceeds $\sim 500\rho_\rms$. 

In the case of the tNFW halo, this happens when $t/t_0 \sim 420$. The vertical arrow at $t/t_0 = 40$ indicates the time of maximum core and marks the snapshot that we use as the initial conditions for the cNFW halo in the numerical experiments described in Section~\ref{sec:DFtNFW} below. In the case of the \Core halo, the halo almost immediately starts to collapse after a very brief period in which it thermalizes its core. The core collapse progresses much faster than for the tNFW halo, reaching core collapse by $t/t_0 \sim 95$. This is a consequence of the much steeper gradient in velocity dispersion (cf. Fig.~\ref{fig:initprofs}), which results in a larger conductive flux outward. 
\begin{figure*}
\centering
\includegraphics[width=0.95\textwidth]{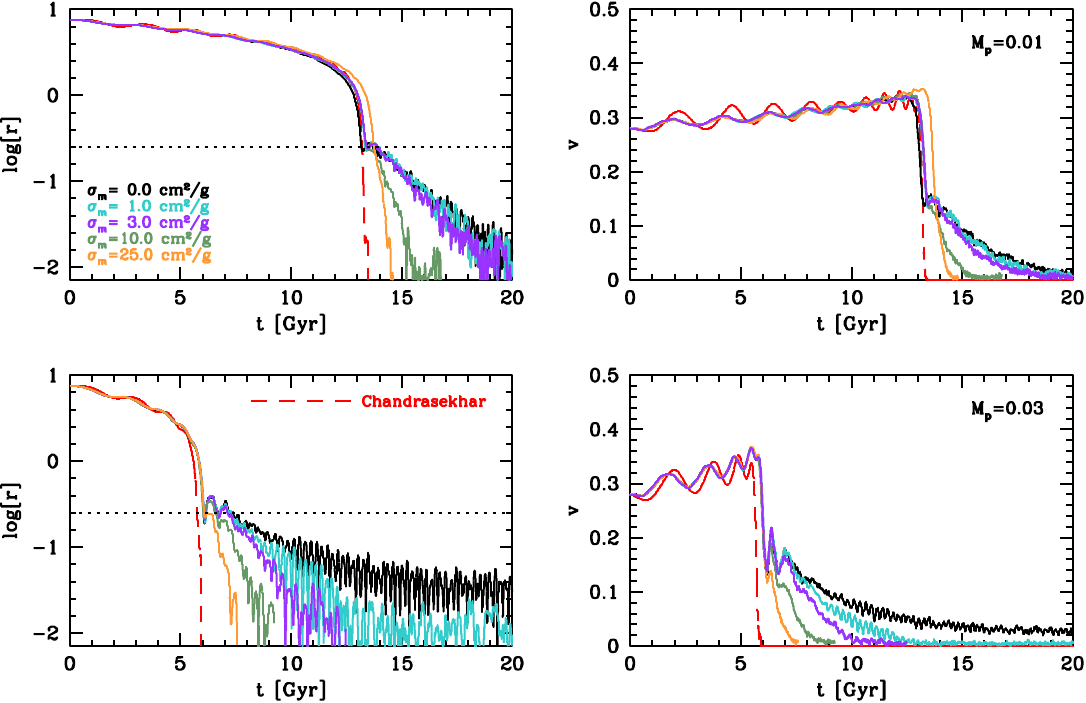}
  \caption{Dynamical friction acting on perturbers of mass $\Mp=0.01$ (top panels) and $\Mp = 0.03$ (bottom panels) in the cNFW halo with different levels of self-interactions, as indicated. Left and right-hand panels show the evolution of the distance and velocity of the perturber with respect to the center-of-mass of the host halo. The red dashed lines indicates the predictions based on Chandrasekhar's dynamical friction formula and the initial density profile of the host halo. The horizontal dotted line marks the stalling radius, $r_{\rm stall}$. Radii and velocities are in model units (see Section~\ref{sec:hosts}).}
\label{fig:cNFW_dynfric}
\end{figure*}

The middle panels of Fig.~\ref{fig:tNFWevol} show the evolution of the total energy of the system. As is apparent, energy conservation in our code remains at the level of $10^{-4}$ ($10^{-3.5}$) for the tNFW (\Core) halo. There is some indication that $|\Delta E/E_0|$ increases during the final stages of core collapse, but at all times it remains below $10^{-3}$. This level of energy conservation is comparable to or better than that of other SIDM simulations \citep[][]{Zhong.etal.23, Palubski.etal.24, Mace.etal.24, Fischer.etal.24, Fischer.etal.25}.

Finally, the right-hand panels of Fig.~\ref{fig:tNFWevol} plot the ratio $\Gamma_{\rm sim}/\Gamma_{\rm pred}$ as a function of time, showing that the scattering rate in the simulation is, at all times, in excellent agreement with the predictions. Together with excellent agreement with the results from the gravothermal fluid code, we conclude that our implementation of self-interactions is accurate. This is further coroborated by Fig.~\ref{fig:tNFWprofs}, which shows snapshots of the density and velocity dispersion profiles of the evolving tNFW halo at four epochs, as indicated. For comparison, the red curves show the predictions from the gravothermal fluid equations and are once again in excellent agreement with the simulation results. Note that the results at $t/t_0=410$ shown in the right-most column correspond to a time well into core collapse. 
\begin{figure*}
\centering
\includegraphics[width=0.95\textwidth]{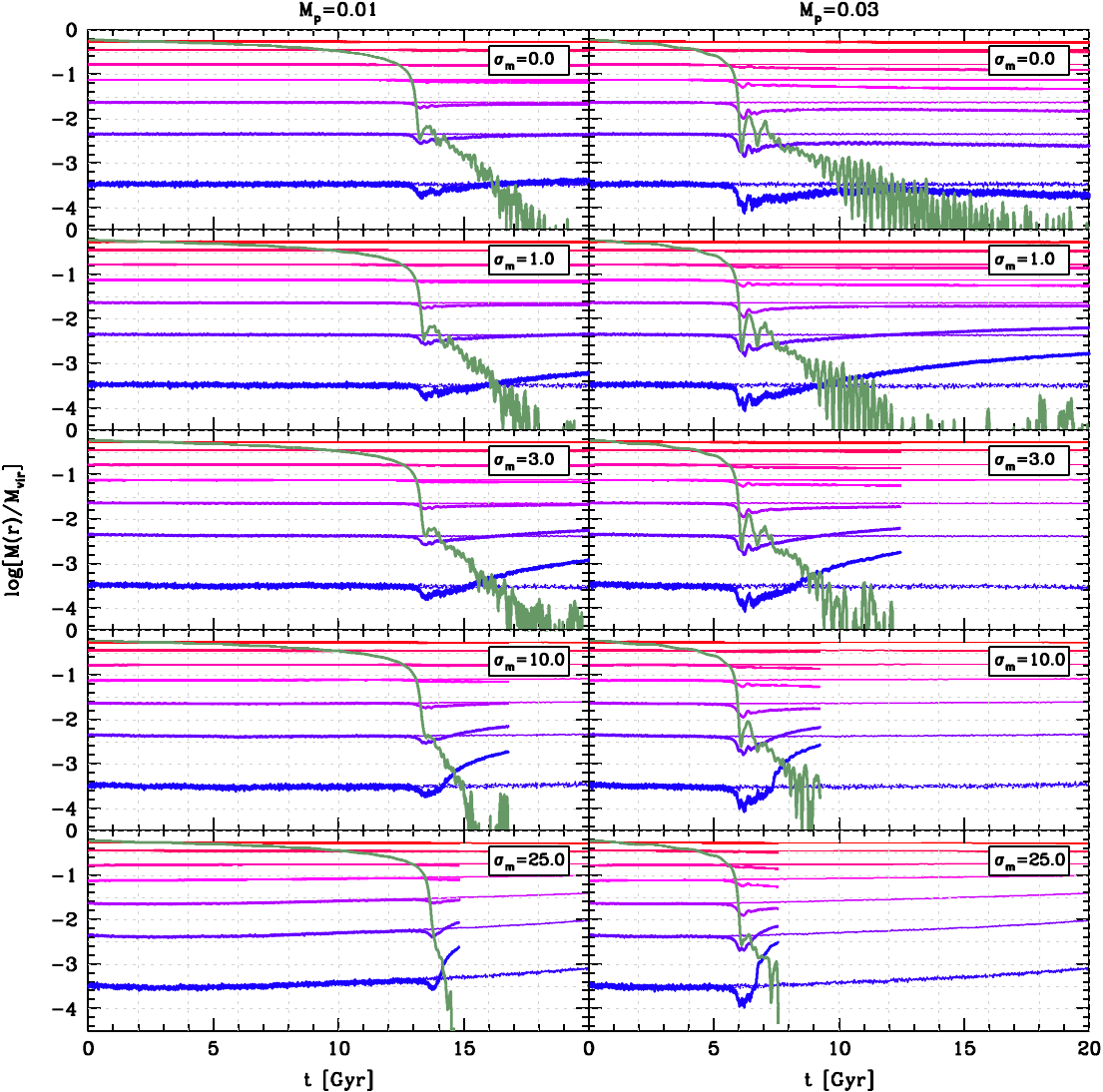}
  \caption{Enclosed masses as a function of time. The colored lines, ranging from blue to red, indicate the evolution of the enclosed mass fractions at different radii: $r/r_\rms = 0.1, 0.25, 0.5, 1.0, 2.0, 5.0$ and $10.0$. Thicker and thinner lines of the same color correspond to the simulations with and without a perturber, respectively. The thick, green line indicates the mass enclosed by the instantaneous radius of the perturber, which declines as the perturber sinks towards the center. Different columns and rows correspond to different perturber masses and collision cross sections, respectively, as indicated. Note how the perturber affects the mass distribution of the host, and catalyzes core collapse in the case of SIDM. The latter is evident from the fact that when the perturber crosses a particular radius, the enclosed mass within that radius (which is indicated by the thick lines and which does not include the perturber itself) increases rapidly compared to the case without a perturber (thin lines).}
\label{fig:cNFWenclosed}
\end{figure*}

\section{Core Stalling}
\label{sec:stalling} 

In order to study whether and how core stalling operates in SIDM halos, we now inject a massive perturber of mass $M_\rmp$ into the cNFW and \Core halos on a circular orbit at an initial radius $r_\rmi$. We have verified (see also \citetalias{Dattathri.etal.25b}) that the value of this starting radius does not have an impact on any of our results, indicating that our results are not affected by potential transients due to the instantaneous introduction of the perturber. We describe the results for the cNFW and \Core halos in turn, starting with the former.

\subsection{Stalling in the cNFW halo}
\label{sec:DFtNFW}

Fig.~\ref{fig:cNFW_dynfric} shows the evolution of the distance and velocity with respect to the halo's center-of-mass of perturbers of mass $\Mp = 0.01$ (top panels) and $\Mp = 0.03$ (bottom panels), in model units. Both perturbers are started from a circular orbit at $r_\rmi = 7.5 r_\rms$ and are modeled as point particles but with a Plummer softening length of $\epsilon_\rmP = 0.03 r_\rms$ and $\epsilon_\rmP = 0.05 r_\rms$, respectively\footnote{Changing the softening length of the perturber has a weak effect on the infall rate, but qualitatively the results are unaffected by changes in $\epsilon_\rmP$}. In the case without self-interactions, indicated by the black lines, the perturber rapidly sinks to $r \sim 0.25 r_\rms$ (indicated by the horizontal dotted line), after which the rate of infall suddenly slows down dramatically. For comparison, the red dashed lines indicate the predictions based on Chandraskehar's formula \citep[][]{Chandrasekhar.43}, obtained by integrating the orbit of the perturber in the (initial) density distribution of the host halo accounting for a dynamical friction 
\begin{equation}
{\rmd \vect{v}_\rmp \over \rmd t} = -4 \pi {G^2 \Mp \over \vp^2} \, \ln\Lambda \, \rho(<\vp) \, {\vect{v}_\rmp \over \vp}\,.
\end{equation}
Here, $\vect{v}_\rmp$ is the velocity of the perturber and $\rho(<\vp)$ is the density of host halo particles with speeds less than $\vp = |\vect{v}_\rmp|$, which is computed assuming that the halo particles locally follow a Maxwellian velocity distribution \citep[see][]{Binney.Tremaine.08}. As is commonly done, we tune the Coulomb logarithm, $\ln\Lambda$, which is rather uncertain \citep[see discussion in][]{MBW10, Just.etal.11}, to fit the simulation results at early times, which yields values in the range $2.5 \lta \ln\Lambda \lta 4$.  The Chandraskehar-based predictions agree with the simulation results\footnote{The small wiggles in the temporal evolution of both radius and (more pronounced) velocity are due to the fact that the orbit of the perturber develops a non-zero eccentricity.}, but only up to the point where the perturber crosses the horizontal dotted line. This is a manifestation of core stalling and in what follows we refer to the radius where the simulation results suddenly and drastically deviate from the Chandrasekhar prediction as the stalling radius, $r_{\rm stall}$. Note that the stalling is not permanent; after $\sim 2 \Gyr$ the perturber continues to sink towards the center of the halo, but at a drastically reduced infall rate. 
\begin{figure*}
\centering
\includegraphics[width=0.95\textwidth]{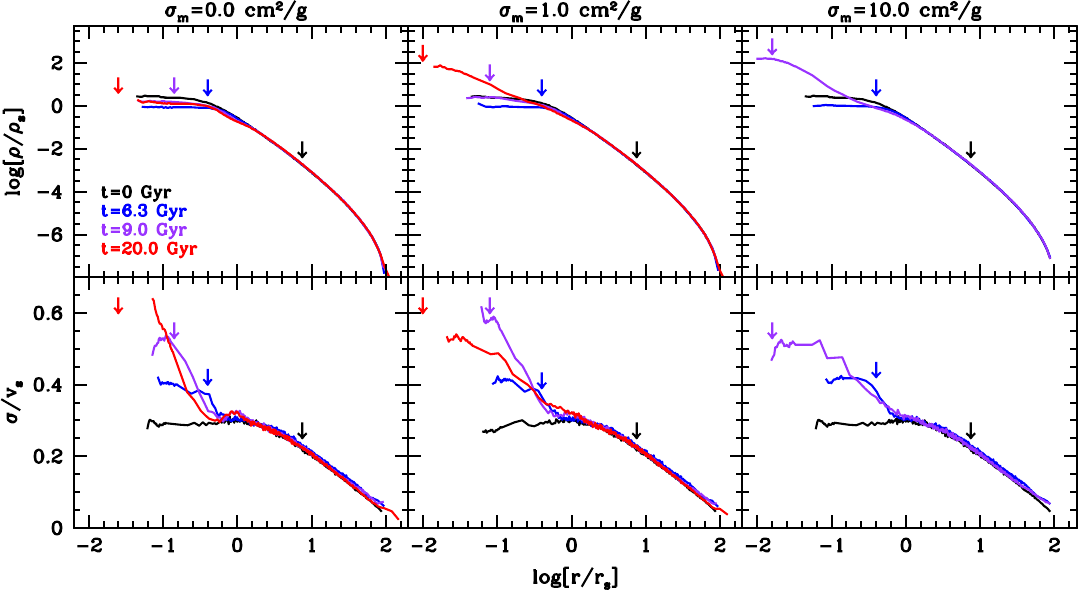}
  \caption{The density (top) and radial velocity dispersion (bottom) profiles for the cNFW halo with a perturber of mass $\Mp = 0.03$ starting at $t=0$ on a circular orbit at $r = 7.5 r_\rms$ (cf. Fig.~\ref{fig:cNFW_dynfric}). Results are shown at four different epochs (different colors, as indicated), and for three different values of the collisional cross section (different columns, as indicated at the top). The arrows mark the halocentric distance of the perturber at the time corresponding to its color. In the case of $\sigma_\rmm = 10\cmg$, the simulation is terminated shortly after $t=9\Gyr$, when the halo has undergone core collapse, which explains why there are no results shown for $t=20\Gyr$. Note that the innermost data point always corresponds to the radius that encloses 30 particles, which thus is smaller when the central density is higher. See text for discussion.}
\label{fig:cNFWdensprof}
\end{figure*}

The colored curves show the same results, but for SIDM simulations with different values for $\sigma_\rmm$, as indicated. In the case of $\Mp = 0.01$, the infall of the perturber is very similar to that in the CDM case and with little dependence on the value of $\sigma_\rmm$ (a consequence of the small scattering rates at those large radii), but only up to the point where the perturber experiences core stalling in the CDM case, at $t \sim 13\Gyr$. Beyond that point in time, there is a pronounced dependence on the interaction cross section: for $\sigma_\rmm \lta 3\cmg$, the results are virtually indistinguishable from those in the case of CDM. However, for $\sigma_\rmm = 10 \cmg$ the post-stalling infall rate is substantially faster. For $\sigma_\rmm = 25 \cmg$, there is no longer any sign of core stalling; while it now takes a little longer for the perturber to reach $r_{\rm stall}$, the perturber sinks past this radius without delay. In the case of $\Mp = 0.03$, the trends are similar; the main difference being that it now takes roughly half as long to reach the core stalling radius\footnote{Note that this time scale does not scale linearly with the inverse of $\Mp$ due to the fact that the perturbers have different softening lengths.}. Note how once again a larger interaction cross section causes a weaker stalling and/or a faster infall inside of the stalling radius. In the case where $\sigma_\rmm = 25\cmg$, there is barely any core stalling at all; except for a minor hiccup in the infall rate around $6 \Gyr$, the perturber continues to sink in rapidly. 

\subsubsection{Impact on the host system}

Fig.~\ref{fig:cNFWenclosed} shows how the infalling perturber impacts the density profile of the host halo. In each panel, the colored lines that range from blue to violet to red indicate the evolution of the halo mass fractions enclosed within the fixed radii of $r/r_\rms = 0.1, 0.25, 0.5, 1.0, 2.0, 5.0$ and $10.0$. The thicker and thinner lines of the same color correspond to simulations with and without a perturber, respectively. Different rows correspond to different collisional cross sections, as indicated (in units of $\cmg$), while the left- and right-hand panels show the results for $\Mp = 0.01$ and $0.03$, respectively. Finally, the thick green line indicates the mass enclosed by the instantaneous radius of the perturber, which declines as the perturber sinks towards the center.  

Let us first focus on the results for $\Mp = 0.03$. In the CDM case (top right panel), once the perturber reaches its stalling radius at $t\sim 6\Gyr$, there is a sudden change in the enclosed mass profile of the host halo. More specifically, the enclosed mass drops and the effect is more pronounced at smaller radii. As discussed below, this corresponds to the perturber tidally perturbing the central core. The energy required to displace the dark matter from the central core outwards is taken from the kinetic energy of the perturber, which explains why, at the same time, the velocity of the perturber drops precipitously (cf. right-hand panel of Fig.~\ref{fig:cNFW_dynfric}). The perturber briefly `stalls' for a period of $\sim 2 \Gyr$, but then continues its inward trajectory, slowly sinking towards the center of its host. During this period, the enclosed density within the innermost radius first increases slightly as a result of adiabatic contraction of the core in response to sinking in of the perturber. At late times, once the perturber has sunk inside of $0.1 r_\rms$, its continued motion through the central core slightly heats the core, causing the enclosed density within the innermost radius to decrease again slightly. 

In the case of SIDM, the behavior is very similar up to $t\simeq 6\Gyr$ when the perturber reaches the core stalling radius.  As in the case of CDM, the perturber initially reduces the centrally enclosed mass, but subsequently the mass enclosed inside $\sim 0.25 r_\rms$ starts to {\it increase} at a rate that is a strong function of $\sigma_\rmm$. We emphasize that this enclosed mass does not include the mass of the perturber. For $\sigma_\rmm = 25 \cmg$ the mass enclosed with $0.1 r_\rms$ increases by more than an order of magnitude in less than 1 Gyr: the perturber, aided by the self-interactions, triggers a sudden and pronounced collapse of the central core. Note how this core collapse goes hand in hand with (i.e., occurs on the same time scale as) the perturber sinking towards the center. 

For the less massive perturber with $\Mp = 0.01$ shown in the left-hand panels, the behavior is qualitatively similar. Once the perturber reaches the core stalling radius, which takes significantly longer than in the case of the more massive perturber, it distorts the central density, after which it triggers a rapid collapse of the core at a rate dictated by $\sigma_\rmm$. The abruptness of the core collapse is most apparent when one compares the results with those in the absence of a perturber shown as the thin lines.
\begin{figure*}
\centering
\includegraphics[width=\textwidth]{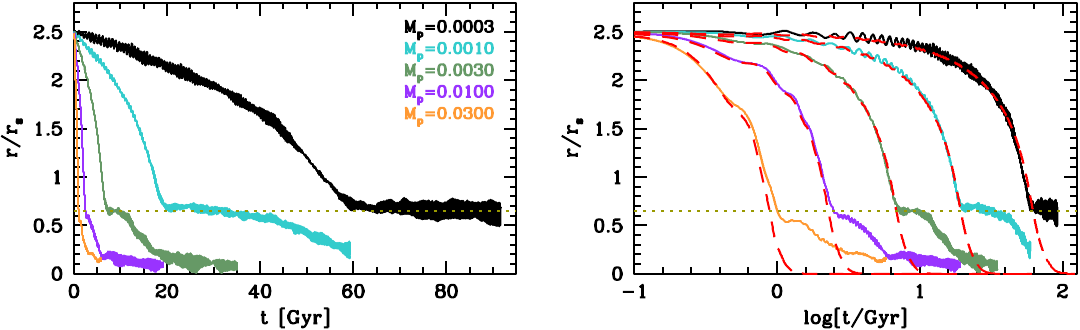}
  \caption{Evolution of the orbital radius of a massive perturber of mass $M_\rmp$ sinking inside the \Core halo assuming CDM. Different colors correspond to different $M_\rmp$, as indicated. Left and right hand panels show results as function of linear and logarithmic time, to better highlight features at late and early times, respectively. Low mass perturbers experience clear stalling, while more massive halos seem to `crunch' through the stalling radius. Note, though, that even the most massive perturber experiences a small `hick-up' sinking towards the center close to the stalling radius, $r_{\rm stall}$, indicated by the horizontal dotted line. The red dashed lines in the right hand panel indicate the predictions based on Chandrasekhar’s dynamical friction formula and the initial density profile of the host halo. These fit the data perfectly, but only up to the point of core stalling.}
\label{fig:coreMsat}
\end{figure*}

Fig.~\ref{fig:cNFWdensprof} is another illustration of the same. It shows the density (top panels) and velocity dispersion (bottom panels) profiles of the cNFW halo with the $\Mp = 0.03$ perturber at four different epochs, as indicated. Different columns correspond to different values of the collisional cross section, while the halo-centric radius of the perturber at each epoch is indicated by a downward-pointing arrow. The left-hand column shows the results for CDM. Note how the perturber initially reduces the central density, enlarging the core, after which the central density slowly creeps up as the perturber sinks inward.  The transition of the central density profile around $t = 6\Gyr$ is extremely rapid. For the first $\sim 5.8\Gyr$, as the perturber sinks from its initial radius $r_\rmi = 7.5$ to $r \sim 0.25$, no significant change in the density profile of the host halo is noticeable. Then suddenly the perturber's velocity plummets from $v \simeq 0.34$ to $0.15$ in less than $0.3\Gyr$. The momentum lost by the perturber is transferred to the central core, which causes a reduction of the central density by roughly a factor of four. This sudden and drastic transfer of momentum is associated with the tidal shredding of the core by the perturber \citep[][]{Goerdt.etal.10}. Subsequently, the perturber temporarily stalls its infall for a period of $\sim 2\Gyr$ before it continues to slowly sink towards the center. During this second inspiral phase, the central density slowly increases again.  By $t = 20 \Gyr$, the central density has roughly doubled compared to what it was during the stalling phase, while the central velocity dispersion has increased dramatically. The evolution during this phase is a complicated competition between two effects: adiabatic contraction and dynamical heating. The sinking of the perturber deepens the central potential, and since the inspiral time scale is long compared to the local dynamical time, this process conserves adiabatic invariants \citep[][]{Blumenthal.etal.86, Weinberg.94a}. This tends to increase both the central density and the velocity dispersion. At the same time, though, the perturber heats the core by transferring its momentum to the dark matter particles. Such dynamical heating typically causes a decrease in the density and, because of the negative heat capacity, a decrease in the central velocity dispersion as well. However, when adding both energy and mass, the latter prevents the core from expanding. This causes the excess heat injected due to dynamical friction to boost the central velocity dispersion over and above the increase that would result purely from adiabatically growing the mass at the center. Due to the collisionless nature of the dark matter, this injected heat cannot be transferred outwards, which explains the dramatic increase in the central velocity dispersion of the host seen in the bottom-left panel.

In the case with $\sigma_\rmm = 1.0\cmg$ shown in the middle column, the density and velocity dispersion profiles are similar to those in the case of CDM up to $t \sim 9\Gyr$. However, from 9 to 20 Gyrs the evolution of the host halo is very different. The perturber catalyzes core collapse, causing an increase in the central density. As discussed in Section~\ref{sec:catalyst}, this catalyzation is driven by the adiabatic contraction and heating of the core due to the sinking perturber. This boosts the central density and velocity dispersion, which in turn increases the outward conduction of heat, thereby further promoting the contraction of the core. And since heat conduction is more efficient for larger cross sections, while stalling is less efficient, the entire process of infall and core collapse proceeds faster for larger $\sigma_\rmm$. Indeed, in the case where $\sigma_\rmm = 10\cmg$, shown in the right-hand column, the system has already completed core collapse by $t=9\Gyr$ (which is why no results are shown for $t=20\Gyr$). Note that in the case of SIDM, the central velocity dispersion ends up lower than in the case of CDM, despite the fact that the core has undergone collapse. This indicated that the self-interactions are extremely efficient in transporting the heat deposited into the core region by the inspiraling perturber outward (see Section~\ref{sec:catalyst} for a detailed discussion).

\subsection{Stalling in the \Core Halo}
\label{sec:DFcore}

Next, we turn our attention to the \Core halo. Fig.~\ref{fig:coreMsat} shows the evolution of the halo-centric distance of perturbers of different mass (different colors, as indicated) that at $t=0$ are injected into the \Core halo along a circular orbit at $r=2.5 r_\rms$. As before, the perturber masses are expressed in model units, that is, in units of the total mass of the ABG halo. Each perturber is modeled as a Plummer sphere with softening length $\epsilon_\rmP = 0.0232$, which is the same value as used for dark matter particles. We have verified that changing the softening length only impacts the time scale at which the perturber sinks in without qualitatively changing the outcome. This simply reflects that more extended objects experience a weaker dynamical friction force \citep[e.g.,][]{Choi.etal.07, Just.etal.11}. Note how the low-mass perturbers experience a pronounced core stalling at $r = r_{\rm stall} \simeq 0.65$  (indicated by the horizontal red dotted line), significantly further out than in the case of the cNFW halo. More massive perturbers appear to stall for a shorter period before they continue their infall to the center. In fact, the most massive perturber, which has a mass that is 3 percent of that of the host halo, barely seems to stall at all. These results are in excellent agreement\footnote{This is noteworthy because the simulations were run with independent $N$-body codes.} with \citetalias{Dattathri.etal.25b}, who showed that this phenomenology is related to evolution in the DF of the host halo. We address this in detail in Section~\ref{sec:discussion} below.

Fig.~\ref{fig:core_dynfric} compares the results for different values of the collision cross section, $\sigma_\rmm$, as indicated. Once more, the perturbers start from a circular orbit at $r=2.5$ and are modeled as Plummer spheres of size $\epsilon_\rmP=0.0232$.  The upper panels show the results for a perturber of mass $\Mp = 0.001$. In the case of CDM ($\sigma_\rmm = 0$), the perturber experiences pronounced and prolonged stalling at $r_{\rm stall} \simeq 0.65$, indicated by the red horizontal dotted line. Similarly to what we found in the case of the cNFW halo, larger values of $\sigma_\rmm$ result in weaker stalling. Even for a modest $\sigma_\rmm = 0.3 \cmg$, the pronounced stalling that is apparent in the case of CDM is gone, and the perturber sinks past $r_{\rm stall}$ without delay. For $\sigma_\rmm = 1.0 \cmg$, the perturber initially sinks past $r_{\rm stall}$ at a rate that is significantly faster than in the case with $\sigma_\rmm = 0.3 \cmg$, but then seems to temporally stall at $r \simeq 0.2$, before continuing its inward trajectory. For $\sigma_\rmm = 3.0 \cmg$, the initial inspiral rate is slower than in the case of CDM. However, there is no indication of any stalling; the perturber passes $r_{\rm stall}$ without interruption and comes to rest at the center of the halo after $\sim 29\Gyr$. Finally, for $\sigma_\rmm = 10 \cmg$, results are only shown up to $t \simeq 12.9 \Gyr$. The reason is that by this time the ABG halo has already undergone core collapse. Interestingly, by $t=12.9 \Gyr$, the perturber has clearly sunk less far than in the other cases, suggesting that it experienced significantly reduced friction. As we will see in Section~\ref{sec:DFstall}, this is due to the fact that the self-interactions have significantly modified the DF. 
\begin{figure*}
\centering
\includegraphics[width=0.95\textwidth]{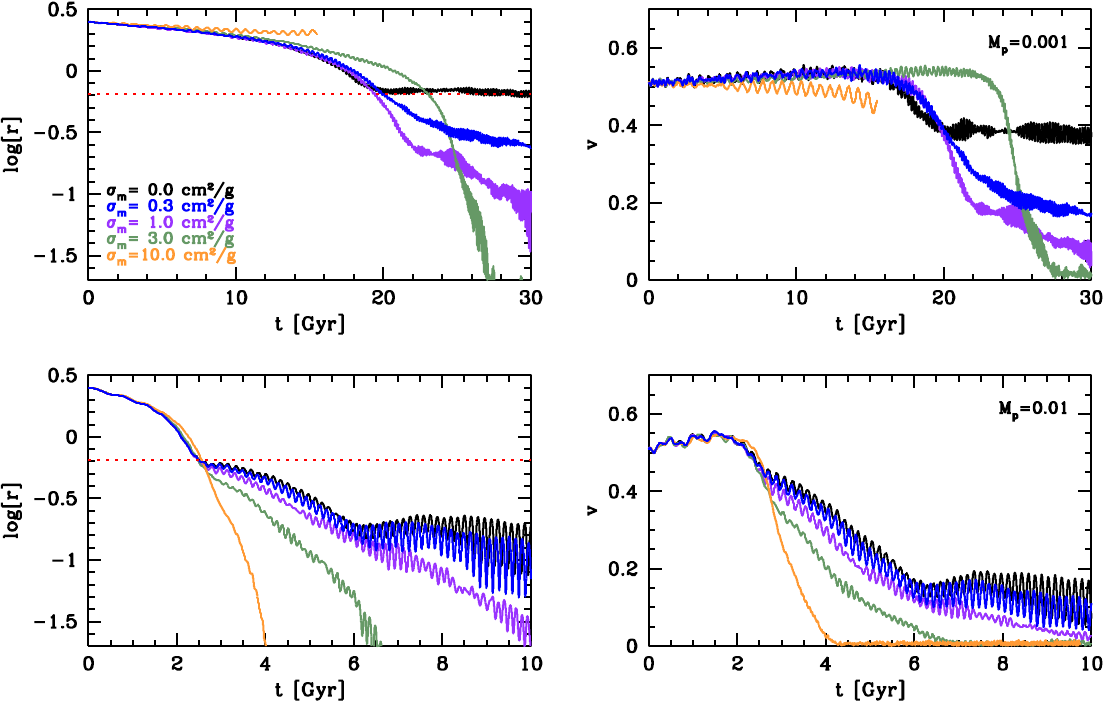}
  \caption{Same as Fig.~\ref{fig:cNFW_dynfric} but for the \Core host halo and a perturber of mass $\Mp = 0.001$ (top panels) and $\Mp = 0.01$ (bottom panels). Note how increasing the cross section for self-interactions suppresses core stalling at late times, while delaying the infall at early times.}
\label{fig:core_dynfric}
\end{figure*}

The bottom panels show the results for a perturber that is ten times more massive. With CDM, the perturber first stalls at $r = r_{\rm stall}$, but only for $\sim 1\Gyr$. Thereafter, it continues to sink inwards, only to stall once more at $t \sim 6.5 \Gyr$ at a radius $r \simeq 0.2$ . When $\sigma_\rmm \gta 1.0 \cmg$ this second stalling disappears and for $\sigma_\rmm \gta 10 \cmg$ there is no longer any sign of stalling.  In fact, for $\sigma_\rmm = 10 \cmg$, the perturber comes to rest at the center of the halo after only $4 \Gyr$. For comparison, with CDM the perturber is still orbiting through the central region of the core at an appreciable speed even after $10 \Gyr$ of evolution. 

\begin{figure*}
\centering
\includegraphics[width=0.95\textwidth]{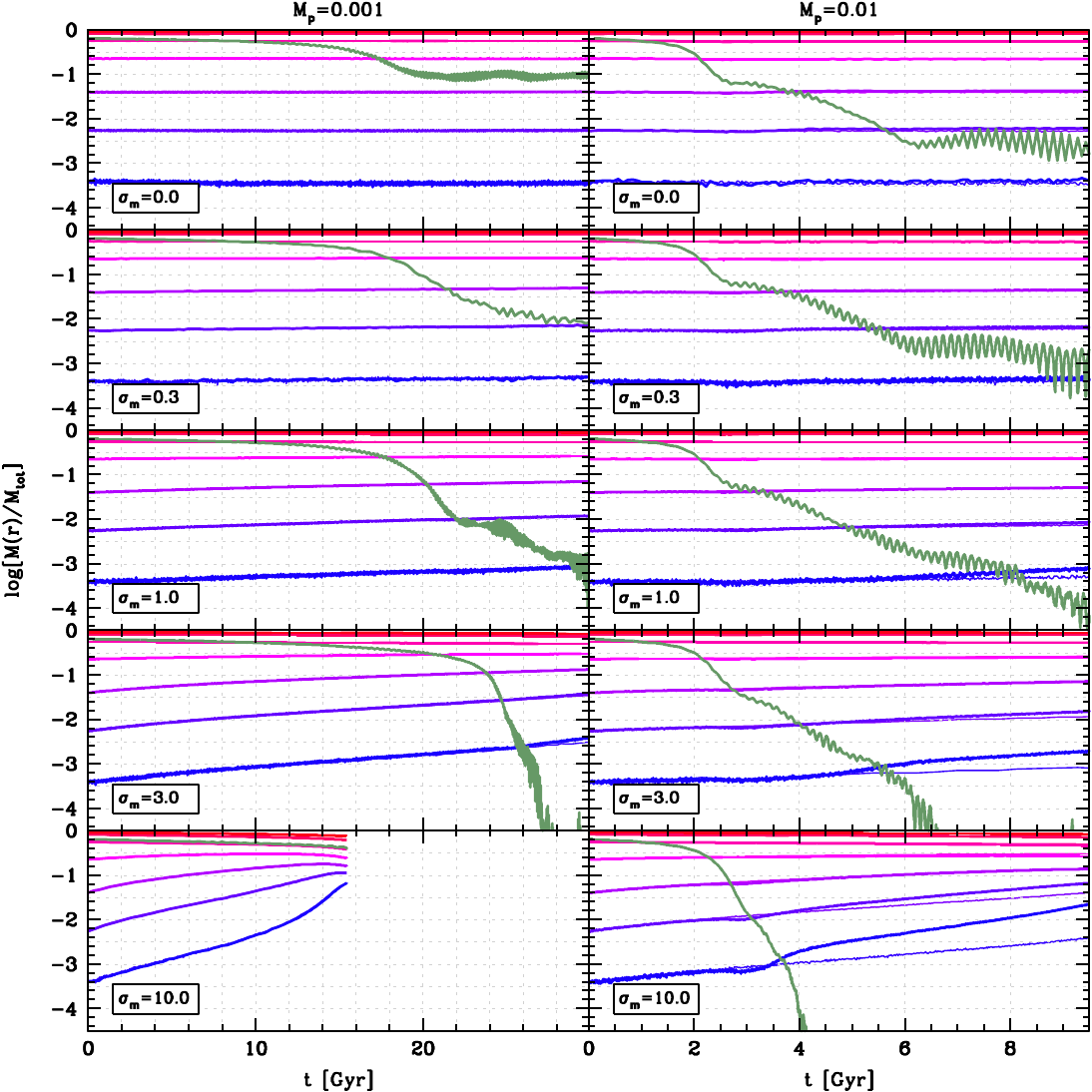}
  \caption{Same as Fig.~\ref{fig:cNFWenclosed}, but for the \Core halo. Note how with increasing self-interaction cross section the perturber sinks more rapidly towards the center, catalyzing core collapse along the way.  In the lower left-hand panel the simulation is terminated at $t = 15.3\Gyr$ when the host halo has undergone core collapse. By that time the perturber has only experienced a small amount of dynamical friction (cf. orange curves in top panels of Fig.~\ref{fig:core_dynfric}).}
\label{fig:COREenclosed}
\end{figure*}

Fig.~\ref{fig:COREenclosed} is similar to Fig.~\ref{fig:cNFWenclosed}, but shows the results of the simulations with the \Core halo. In the case with $\Mp = 0.001$, shown in the left-hand column, the perturber has barely any impact on the host halo. This is evident from the fact that the thin colored lines that show the results without a perturber are barely if at all visible; they lie below the thick colored lines. There are two clear trends with increasing cross section for self-interactions: (i) the perturber experiences less stalling (i.e., it sinks faster after it has crossed the stalling radius), and (ii) the halo experiences more rapid core collapse. In the case with $\sigma_\rmm = 10 \cmg$, shown in the bottom panel, the core of the host halo has collapsed by $t \sim 15\Gyr$, when we are forced to terminate the simulation. Up to that point, the perturber has barely sunk in at all and has had no noticeable impact on the host system (i.e., without perturber the host halo experiences core collapse in the same way and in the same amount of time). For $\Mp = 0.01$, shown in the right-hand column, the sinking rate of the perturber is significantly enhanced, as is the impact that the perturber has on its host halo. In particular, for $\sigma_\rmm \gta 1.0 \cmg$ it is evident that the perturber causes an acceleration of the core collapse of the host halo, albeit less pronounced as in the case of the cNFW halo. All in all, though, the trends are qualitatively similar for the two different host halos.

However, there is one subtle but interesting difference. In the cNFW case, the sinking perturber causes a sudden change in the enclosed mass profile of the host system that we associated with the perturber tidally perturbing the central core. Such a sudden and dramatic change of the host system is absent in the case of the \Core halo. As we discuss in Section~\ref{sec:DFstall}, this relates to the (initial) shape of the host halo's distribution function.

\section{Dynamical Buoyancy and the Dipole Instability}
\label{sec:dipole}

We now turn our attention to dynamical buoyancy and the associated dipole instability. Here, to study the impact of self-interactions, we consider the \Cusp halo, which in the case of CDM is known to manifest a pronounced dipole instability \citep[][]{Dattathri.etal.25a}. We evolve the halo in isolation and evolve it for roughly a Hubble time for four different values of $\sigma_\rmm$: 0 (CDM), $0.1$, $0.3$, and $1.0 \cmg$. In order to be able to probe the impact of dynamical buoyancy, we run two simulations for each value of $\sigma_\rmm$; one without a perturber and one in which, at $t=0$, we place a perturber of mass $M_\rmp = 0.01$ at rest at the center of the system. In the latter case, the DF that is used to setup the ICs is adjusted accordingly\footnote{We find that simply placing the perturber at rest in the simulation that is initialized without perturber yields results that are virtually indistinguishable.}. The perturber particle has the same softening length as the regular particles but is not susceptible to self-interactions. It represents a massive object such as a SMBH or a nuclear star cluster that, in the absence of buoyancy and stalling, will have a tendency to sink towards the center by dynamical friction. 

\begin{figure*}
\centering
\includegraphics[width=\textwidth]{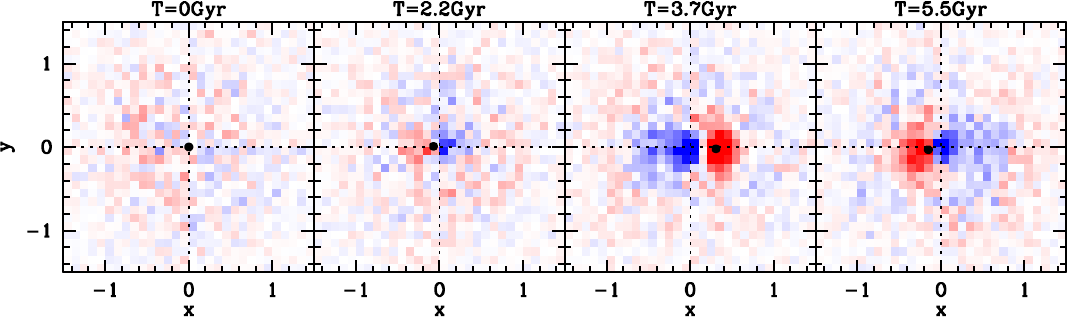}
  \caption{Four snapshots of the excess surface density $\Delta \Sigma = \Sigma(x,y) - \Sigma_0(x,y)$ of the \Cusp system evolved assuming CDM (no self-interactions), where $\Sigma_0(x,y)$ is the analytical surface density at $t=0$. Red and blue pixels indicate over- and under-densities respectively \citep[cf. Fig.~3 in][]{Dattathri.etal.25a}. At $t=0$ the system starts out symmetrically, with only some discreteness noise. By $t=1.8 \Gyr$, a small dipole appears above the noise, which grows in amplitude as it rotates and sloshes through the central region. At each snapshot the solid dot marks the center-of-mass of the 500 densest particles, the trajectory of which is shown by the blue curve in the top-right panel of Fig.~\ref{fig:dipole}.}
\label{fig:dipole_evol}
\end{figure*}

For completeness, in order to demonstrate the presence of the dipole mode, Fig.~\ref{fig:dipole_evol} shows the projected over/underdensity, $\Delta \Sigma = \Sigma-\Sigma_0$, where $\Sigma_0$ is the unperturbed analytic projected density, for four snapshots of the CDM simulation without perturber. While the snapshot at $t=0$ mainly reveals shot noise due to the discreteness of the $N$-body system, at later times a clear dipole is evident. Importantly, the dipole is not stationary but rather rotating in a plane. We emphasize that these results are in excellent agreement, both qualitatively and quantitatively, with those of \citet[][cf. their Fig.~3]{Dattathri.etal.25a}, despite significant differences in the simulations\footnote{The simulation analyzed in \citet{Dattathri.etal.25a} used $N_\rmp = 5 \times 10^6$ particles with a mass spectrum that boosts mass resolution near the center, and was as run using the code {\tt GyrfalcOn} \citep{Dehnen.02} with hierarchical time stepping. For comparison, the simulation presented here only uses $N_\rmp = 10^6$ particles, all of identical mass, and is run using {\tt treecode} without hierarchical time stepping.}. As discussed in detail in \citet{Dattathri.etal.25a}, the strong dipole apparent in the center is not actually the dipole mode; rather, it is the central $r^{-0.5}$ cusp that has been dislodged and set in motion by the mode. The cusp continues to slosh through the central region of the system, along a roughly elliptical orbit, without any sign of ever coming back to rest.

\begin{figure*}
\centering
\includegraphics[width=0.9\textwidth]{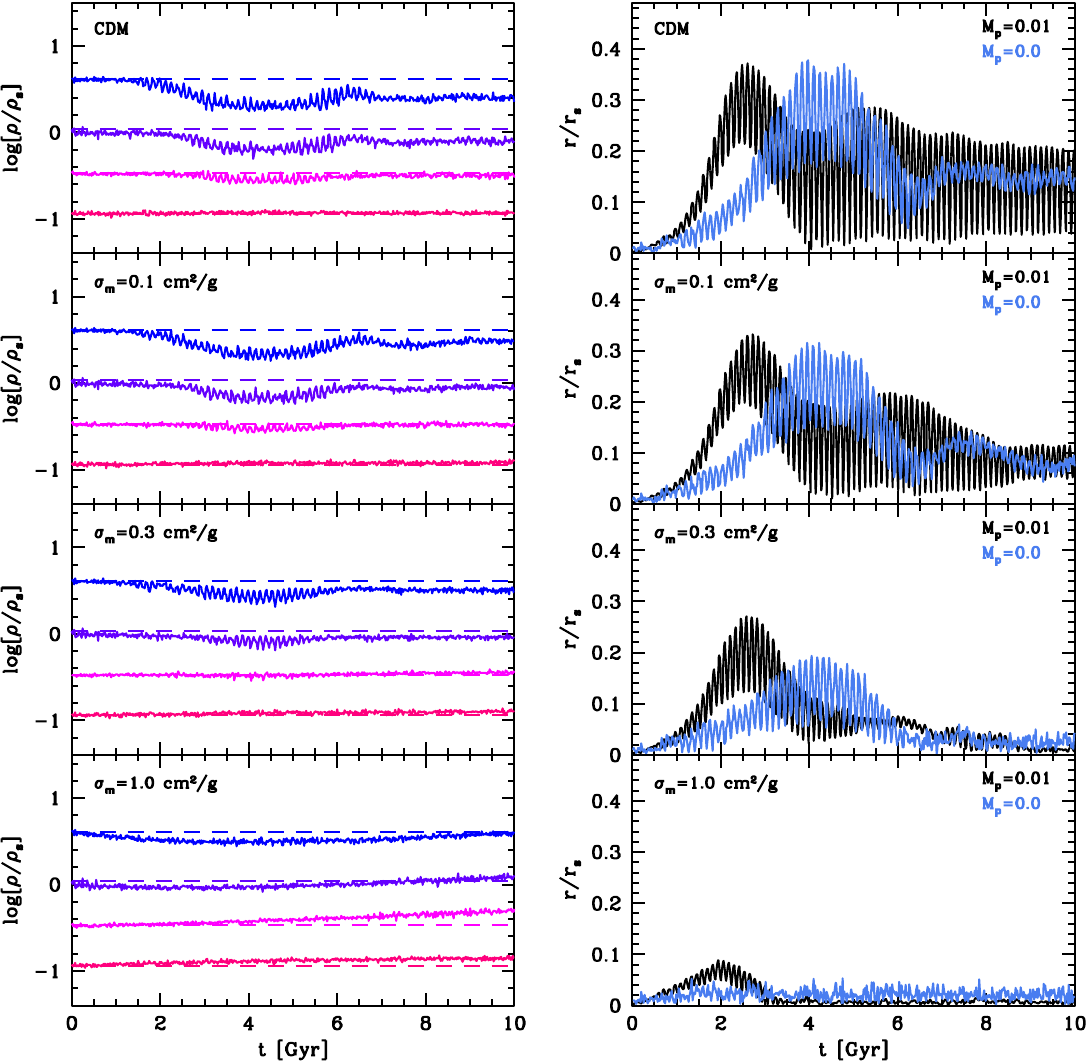}
  \caption{{\it Left-hand panels:} The evolution of the density of the \Cusp halo in simulations without any perturber but with different values for $\sigma_\rmm$ (different rows), as indicated. In each panel, the different colors correspond to the densities measured at different radii; $\log[r/r_\rms]=-1.25$ (blue), $-1.0$ (violet), $-0.75$ (magenta), and $-0.25$ (red). The dashed, horizontal lines of corresponding color indicate the values of the density at $t=0$ and are shown to highlight by how much the densities evolve. The strong evolution in the central densities in the case of CDM (top panel) is due to the dipole instability. Note how the instability becomes weaker with increasing $\sigma_\rmm$; for $\sigma_\rmm = 1.0\cmg$ (red curves), the halo starts to undergo core collapse before the dipole instability can develop. {\it Right-hand panels:} the colored lines indicate the temporal evolution of the cusp in the four simulations shown in the left-hand panels. Note how, in the case of CDM (top panel), the cusp starts moving outward along a highly eccentric orbit, after which it settles at $r \sim 0.15 r_\rms$. As discussed in the text, this reflects the cusp being dislodged, and set in motion, by the dipole mode. This effect becomes weaker with increasing $\sigma_\rmm$ (going from top to bottom). The black lines in each panel indicate the halo-centric radii of a massive perturber ($M_\rmp=0.01$) which at $t=0$ is positioned at rest at the CoM of the system. Note how, for small enough $\sigma_\rmm$, the perturber moves outward (dynamical buoyancy). The motion is similar to that of the cusp in the simulations {\it without} perturber, indicating that dynamical buoyancy is a direct manifestation of the dipole instability. See text for a detailed discussion.}
\label{fig:dipole}
\end{figure*}

The left-hand panels of Fig.~\ref{fig:dipole} show the evolution of the density at four different radii with respect to the system's center of mass (different colors). The dashed lines of the corresponding color indicate the densities at $t=0$ and are shown to highlight the extent to which the densities evolve over time. Different panels correspond to different values of the interaction cross section, $\sigma_\rmm$, as indicated. First, let us focus on the case of CDM (top panel). Clearly, the central densities at $r \lta 0.1 r_\rms$ (blue and violet curves) reveal strong evolution, consisting of a large oscillation with small oscillations superimposed. This reflects the onset and development of the dipole mode, with the small oscillations reflecting the motion of the cusp along its elliptical orbit. Note that most of the action is confined to the central region of the system; the density at $\log[r/r_\rms]=-0.25$ (red curve) remains constant with time within the noise. 

As we dial up the self-interactions, the amplitude of the oscillations becomes less pronounced, and for $\sigma_\rmm = 1.0\cmg$ (bottom panel), the small-scale oscillations have completely disappeared. Rather, after an initial decline, the density in the entire central region starts to increase, which reflects the onset of core collapse. 

The right-hand panels of Fig.~\ref{fig:dipole} show how all this is related to buoyancy. Once again, different panels correspond to different values for the self-interaction cross section, as indicated. In each panel, the black curve shows the temporal evolution of the center of mass of the massive perturber that at $t=0$ starts out at rest at the center of the system. Note how, in the case of CDM, the perturber moves outward during the first $2 \Gyr$, reaching as far as $0.3 r_\rms$; this is a manifestation of dynamical buoyancy. Later, it sinks back somewhat before stalling at $\sim 0.1 r_\rms$ along a highly eccentric orbit. We also computed the center of mass of the 500 densest particles, which accurately traces the motion of the dislodged cusp, and find that it accurately follows the trajectory of the massive perturber (not shown). This indicates that the buoyancy experienced by the massive perturber is a direct manifestation of the dipole instability; the massive perturber simply `rides along', and remains centered on, the dislodged cusp (see also \citetalias{Dattathri.etal.25b}).

The azure curve in the top-right panel of Fig.~\ref{fig:dipole} represents the trajectory of the center of mass of the 500 densest particles (which basically tracks the location of the original cusp) in the simulation {\it without perturber}. As in the simulation with perturber, the dipole instability dislodges the cusp and sets it in motion along a trajectory that is very similar to that of the massive pertuber (cf. Fig.~\ref{fig:dipole_evol}), albeit with a time-lag of $\sim 1.5\Gyr$. This time lag reflects the fact that the massive perturber has slightly modified the DF of the ICs, which in turn impacts the growth of the dipole mode. 

With increasing cross section, the amplitudes of the radial excursions of both the massive perturber (black curves) and the dislodged cusp in the simulations without perturber (azure curves) become weaker and weaker. In fact, for $\sigma_\rmm = 1.0\cmg$ (bottom-left), the massive perturber initially moves slightly outward (never exceeding $0.1 r_\rms$), but then rapidly sinks back and settles at the center of the system. The small remaining `jitter' in the radial position of the massive perturber is due to Brownian motion. Hence, increasing $\sigma_\rmm$ inhibits the growth of the dipole instability and thus also the amplitude of dynamical buoyancy.

\section{Discussion}
\label{sec:discussion}

We have shown that self-interactions among dark matter particles have a profound impact on core dynamics. In particular, self-interactions suppress core stalling, dynamical buoyancy, and the onset of the dipole instability. In fact, for sufficiently large $\sigma_\rmm$ core stalling can be completely avoided, and the perturber sinks to the center much faster than in the case of CDM. Interestingly, in doing so it causes a drastically accelerated core collapse. Here, we discuss the physical origin of these effects. 
\begin{figure*}
\centering
\includegraphics[width=0.94\textwidth]{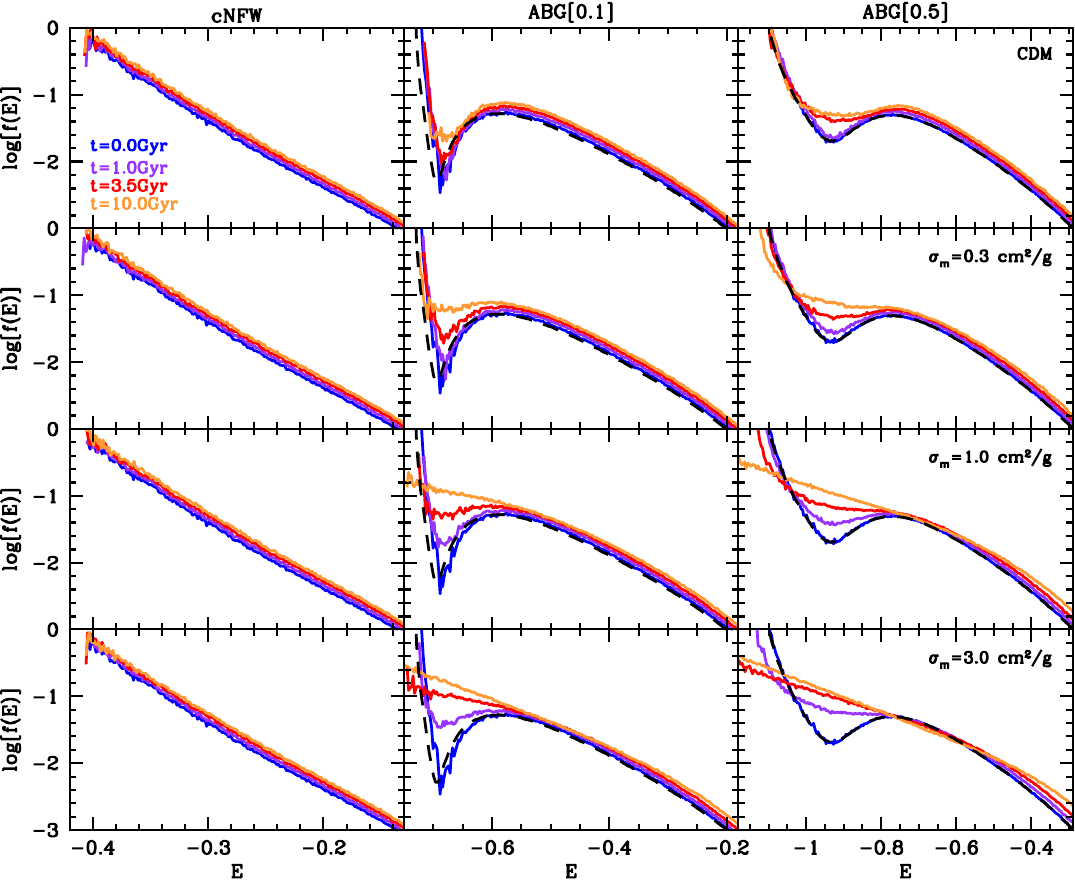}
  \caption{Evolution in the DFs for the three different cored systems studied here; from left to right, these are the cNFW halo, the \Core halo, and the \Cusp halo. Different colored lines correspond to different times, as indicated. Different rows correspond to different collisional cross sections, as indicated in the right-hand panels. Top panels correspond to CDM (i.e., $\sigma_\rmm=0$). The black dashed lines in the two right-most columns indicate the analytical DFs at $t=0$ used to set up the ICs of the halos, and are shown for comparison.}
\label{fig:df_dipole}
\end{figure*}

As discussed in Section~\ref{sec:kinetics}, both core stalling and the dipole instability are related to features in the DF of the host system. In particular, stalling happens when the distribution function of the host system has a plateau (i.e., $\rmd f/\rmd E \simeq 0$) at the orbital energy of the perturber, while systems with an inflection in their DF, that is, a region in energy-space where $\rmd f/\rmd E > 0$, are unstable to the dipole mode\footnote{We emphasize that it remains unproven whether this is always the case. However, in all such systems studied by the authors we have found the system to be unstable.}. Self-interactions modify the DF, and this might thus explain why core dynamics in SIDM halos can be substantially different from that in CDM halo. In particular, self-interactions cause a diffusion in phase space, which tends to erase (small-scale) gradients, and drive the system toward an isothermal form $f(E) \propto \exp(-E/\sigma^2)$ \citep[][]{Spitzer.Shapiro.72}, at least in the central core region. Since the corresponding $\rmd f/\rmd E$ is always negative, we ought to expect that in systems in which the self-interactions have established isothermality, core stalling and/or buoyancy are absent. This is qualitatively consistent with the trends seen in Section~\ref{sec:stalling}. In what follows, we test this hypothesis in detail by computing the evolution in the DF as the perturber sinks towards the center. For each simulation output, we calculate the phase space distribution of the simulation particles as a function of their instantaneous energy, $f(E)$, using the method of \citet{Dattathri.etal.25a}, summarized in Appendix~\ref{sec:compDF}.
\begin{figure}
\centering
\includegraphics[width=0.48\textwidth]{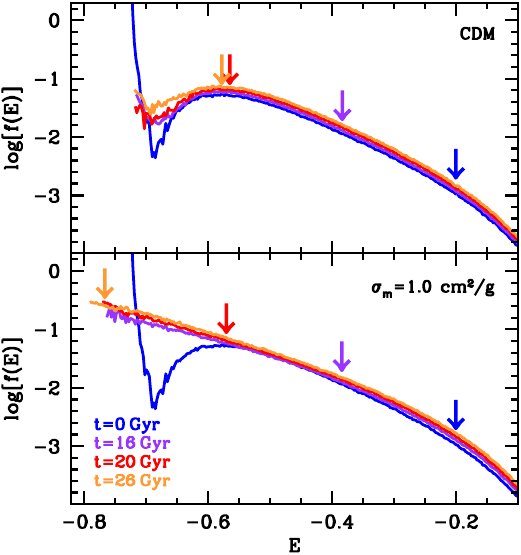}
  \caption{Evolution in the DFs of the \Core halo in the presence of a massive perturber with $M_\rmp = 0.001$ in the case of CDM (top panel) and SIDM with $\sigma_\rmm = 1.0 \cmg$ (bottom panel). Results are shown at four different epochs: $t=0$ (blue), $t=16 \Gyr$ (violet), $t=20 \Gyr$ (red) and $t=26 \Gyr$ (orange). Downward arrows in matching color indicate the orbital energy of the perturber at the corresponding times. For comparison, the radial evolution of the perturber in both cases is shown by the black and violet lines in the top-left panel of Fig.~\ref{fig:core_dynfric}.}
\label{fig:df_pert}
\end{figure}

\subsection{Suppression of the dipole instability \& buoyancy}
\label{sec:dipolebuoyancy}

We first focus on the cNFW halo, the \Core halo, and the \Cusp halo in isolation (i.e., without perturber). Fig.~\ref{fig:df_dipole} shows the DFs of the three halos (different columns) at four different epochs (different colors), which have been slightly offset from each other in the vertical direction for the sake of clarity. Different columns correspond to different values for the cross section for self-interactions, as indicated, with the top row corresponding to CDM (i.e., $\sigma_\rmm =0$). The DF of the cNFW halo, shown in the left-hand panels, is close to a pure exponential and shows little to no evolution over the period of $10\Gyr$ shown here. This is consistent with expectations. After all, the cNFW halo was created by evolving a (truncated) NFW halo with strong self-interactions ($\sigma_\rmm=25\cmg$) until the original cusp was transformed into an isothermal core, which has a DF of the form $f(E) \propto \exp(-E/\sigma^2)$. As mentioned in Section~\ref{sec:hosts}, the characteristic collision time for the cNFW halo is $t_0 = 1.91\Gyr \, (\sigma_\rmm/\cmg)^{-1}$, and core collapse takes on the order of 400 collision times (see top-left panel of Fig.~\ref{fig:tNFWevol}). Hence, for the largest cross section considered here, $\sigma_\rmm = 3\cmg$, core collapse takes over $250\Gyr$. Hence, it is not surprising that the DFs show no sign of evolution over the duration of $10\Gyr$ shown here,

Both the \Core and \Cusp halos have initial DFs that do not decrease monotonically with $E$, but rather reveal a pronounced dip near $E \sim -0.7$, which results in an inflection and thus a region in phase space where $\rmd f/\rmd E > 0$, in violation of Antonov's stability criterion. Consequently, as previously demonstrated in \citet[see also Section~\ref{sec:dipole} above]{Dattathri.etal.25a}, both systems are unstable to the dipole instability. When the dipole mode develops in strength, non-linear effects such as orbit trapping become important. As particles undergo separatrix crossing and become trapped into librating orbits, their actions undergo permanent changes, causing irreversible evolution of the DF. This evolution is evident in the top-right panel, which shows how, in the case of CDM, the dipole mode slowly flattens the dip in the DF of the \Cusp halo. By $t=10\Gyr$, most of the dip in the original DF has disappeared \citep[see][for a detailed discussion]{Dattathri.etal.25a}. In the case of the \Core halo (top middle panel), the dipole mode is significantly weaker and takes longer to become nonlinear. This explains why after $10\Gyr$ the DF has evolved very little, with only a slight reduction in the strength of the inflection.

In the case of SIDM, the self-interactions cause diffusion in phase space that eliminates the local suppression in the DF on a time scale that, as expected, decreases with increasing $\sigma_\rmm$. In both the \Core and \Cusp halos, a cross section of $1 \cmg$ suffices to completely remove the inflection in under $10\Gyr$. Note that the interactions drive the distribution function towards isothermality, which explains why $f(E)$ takes on an exponential form at the most negative $E$ (that is, in the strongly bound central region).

\subsection{Reduced core stalling in SIDM}
\label{sec:DFstall}

In order to demonstrate how core stalling is impacted by the detailed shape of the host halo's DF, we focus on two representative simulations that follow the evolution of a perturber of mass $M_\rmp = 0.001$, starting on a circular orbit at $r=2.5 r_\rms$ in the \Core halo. One simulation assumes CDM ($\sigma_\rmm=0$), while the other assumes SIDM with $\sigma_\rmm= 1.0 \cmg$. The radial evolution of the perturber in both cases is shown in the top left panel of Fig.~\ref{fig:core_dynfric} as the black and violet curves, respectively.  In the case of CDM, the perturber sinks to $r = r_{\rm stall} = 0.65 r_\rms$ in $\sim 20\Gyr$, at which point the radial infall comes to an abrupt end, and the perturber stalls out at $r_{\rm stall}$ indefinitely. In the case of SIDM, the evolution is virtually identical for the first $20\Gyr$; however, rather than stalling at $r_{\rm stall}$, the perturber continues to sink inward, reaching $r \simeq 0.2$ by $t=26\Gyr$.

Fig.~\ref{fig:df_pert} shows the DFs of the \Core halo in these two simulations at four different epochs. The top and bottom panels correspond to the CDM and SIDM simulations, respectively. The initial DF, which is identical for the CDM and SIDM cases, is shown in blue in both panels. As already shown in Fig.~\ref{fig:df_dipole}, this system has a pronounced inflection in its DF at $E \simeq -0.7$, rendering it unstable to the dipole mode. However, as shown above, the instability takes a very long time to develop and therefore only has a marginal impact on the system's evolution over the period considered here. The violet, red, and orange curves show the system's DFs at $t=16\Gyr$, $t=20\Gyr$, and $t=26\Gyr$, respectively, while the downward arrows of the same color indicate the orbital energy of the perturber at those corresponding epochs. Note how in the case of CDM, the red arrow indicates that at $t=20\Gyr$, when stalling starts, the DF at the orbital energy of the perturber has a vanishing gradient, that is $\rmd f/\rmd E = 0$. This explains why the system experiences core stalling (see Section~\ref{sec:kinetics}). Note that the orbital energy of the perturber at $t=26\Gyr$, indicated by the orange arrow, is almost identical to that of $6 \Gyr$ earlier (red arrow), indicating that the perturber has indeed stalled its inward motion. Note also that from $t=0$ to $t=26\Gyr$ the depth of the dip in the initial DF has reduced significantly. As discussed in \citetalias{Dattathri.etal.25b}, this is the combined effect of the dipole instability and the perturber modifying the DF of the host system.
\begin{figure*}
\centering
\includegraphics[width=0.95\textwidth]{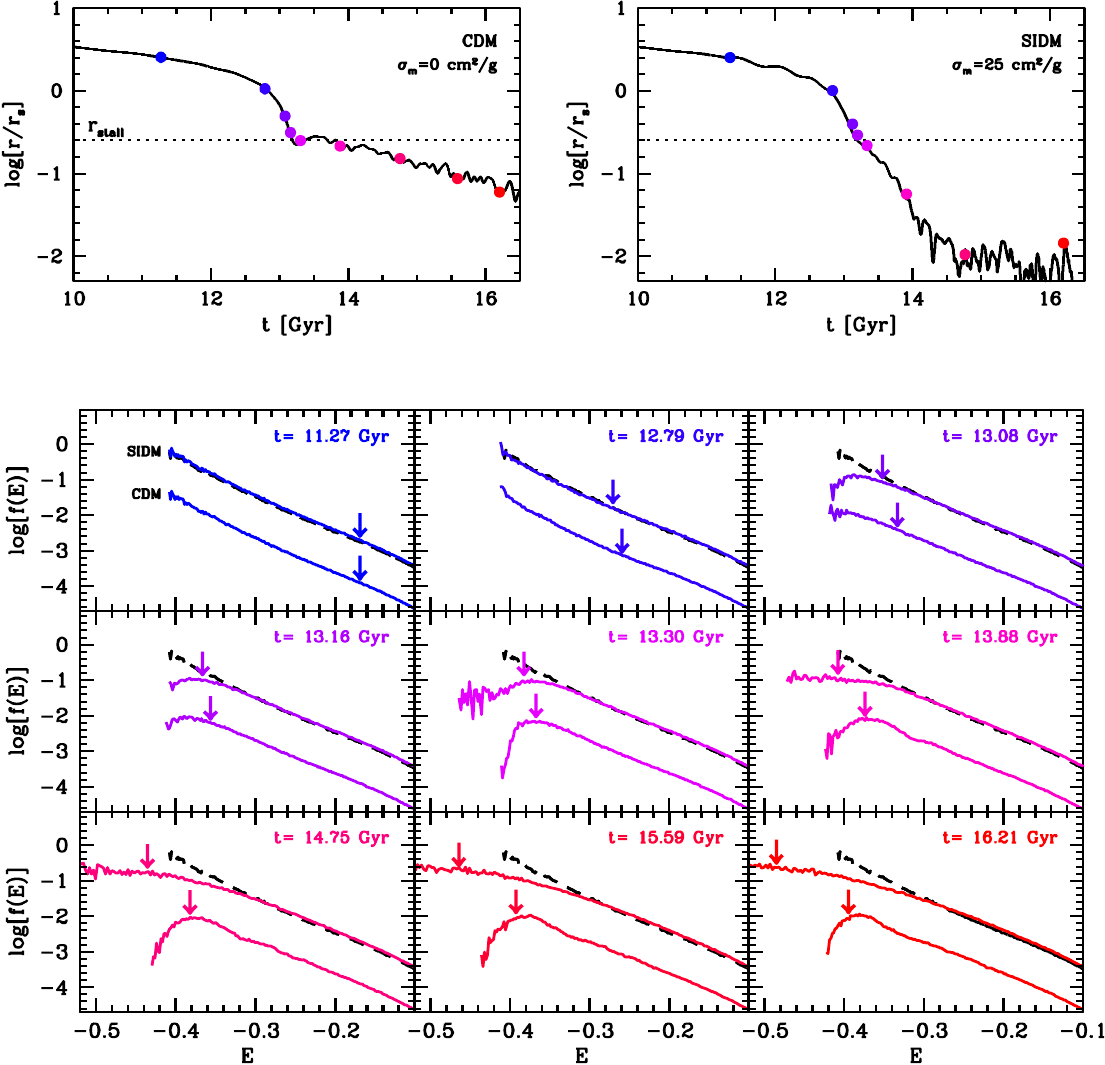}
  \caption{The impact of a perturber of mass $M_\rmp = 0.01$ sinking inside the cNFW halo. Upper panels show the halocentric radius of the perturber as a function of time in the case of CDM (right) and SIDM (left). The lower $3 \times 3$ panels show the corresponding DFs at 9 different epochs, with the downward arrow indicating the energy of the perturber at the corresponding time. The DFs for the CDM case are offset by $-1.2$ dex for clarity. As in Fig.~\ref{fig:df_pert}, the downward pointing arrows indicate the instantaneous orbital energy of the perturber. The black dashed lines indicate the initial DF at $t=0$ and is shown for comparison. Note how in the case of CDM, the perturber introduces an inflection in the DF, resulting in a bump where $\rmd f/\rmd E$ vanishes, which in turn causes the perturber to stall. In the case of SIDM, the self-interactions oppose this by driving the DF towards an isothermal; consequently $\rmd f/\rmd E$ remains negative and the perturber sinks all the way to the center.}
\label{fig:cNFW_DF}
\end{figure*}

In the case of SIDM with $\sigma_\rmm = 1\cmg$, shown in the lower panel, the inflection in the DF has been completely erased, already at $t=16\Gyr$, as a consequence of the self-interactions. By the time the perturber reaches $r_{\rm stall}$ at $t=20 \Gyr$, the gradient in the DF at the orbital energy of the perturber has become negative, as opposed to zero in the case of CDM,  which explains why the perturber continues to sink inwards without stalling. By $t=26 \Gyr$, the perturber has sunk almost entirely to the center of the system, which has started to undergo core collapse; this is evident from the fact that the DF toward the center (at small $E$) has become a featureless exponential consistent with an isothermal core. Note also that the DF at large $E$ becomes somewhat shallower with time. Since the LBK torque is directly proportional to the local gradient in the DF, this explains why, in the case of larger cross sections, the perturber takes longer to sink in (see top-left panel of Fig.~\ref{fig:core_dynfric}).

Finally, to really drive home the point, Fig.~\ref{fig:cNFW_DF} shows the results for a perturber of mass $M_\rmp = 0.01$ sinking inside the cNFW halo. Recall, from Fig.~\ref{fig:df_dipole}, that this halo starts out with a DF that is almost a perfect power law, without any bumps or inflection points. Hence, naive predictions based on the DF gradient are that the perturber should sink to the center without interruption. However, as shown in Fig.~\ref{fig:cNFW_dynfric}, this is not the case. In particular, in the case of CDM the perturber stalls at a stalling radius $r_{\rm stall} \simeq 0.25 r_\rms$. However, in the case of SIDM, as long as the cross section is large enough, the perturber sinks through $r_{\rm stall}$ and sinks rapidly toward the center of the system. The top panels of Fig.~\ref{fig:cNFW_DF} show the evolution of the halocentric distance of the perturber in the case of CDM (upper left) and SIDM with $\sigma_\rmm=25\cmg$ (upper right). These are identical to the corresponding curves in the upper panels of Fig.~\ref{fig:cNFW_dynfric}, except that here we zoom in on the period between 10 and 16.5 Gyr. The solid colored dots indicate specific epochs at which the distribution functions $f(E)$ are shown in the $3\times 3$ grid of panels at the bottom of Fig.~\ref{fig:cNFW_DF} (in corresponding color). Note that we have offset the DF for the CDM case by $-1.2$ dex for the sake of clarity. The downward pointing arrows indicate the instantaneous orbital energy of the perturber. Initially, up until $t \sim 13 \Gyr$ the DF of the host remains largely unchanged with respect to the initial power law, indicated by the black dashed line).  However, once the perturber comes close to $r_{\rm stall}$, it causes a strong distortion of the host (cf. left-hand panels of Fig.~\ref{fig:cNFWdensprof}) that creates a suppression in the DF at $E \lta -0.42$. This results in a bump where $\rmd f/\rmd E = 0$, and this is turn causes the perturber to stall its inward motion. In the case of CDM, the suppression remains, although the DF continues to undergo small changes due to interactions with the perturber. This in turn causes the perturber to slowly drift inwards over the period from 13 to 16.5 Gyr. In the case of SIDM, the self-interactions oppose the creation of an inflection in the DF by continuously driving the system towards isothermality. With a large cross section of $25 \cmg$, the self-interactions are able to re-establish a negative gradient in the DF shortly after the perturber introduced a suppression, which is why the perturber continues to sink inward with only a minor `hick-up' in its sinking rate around $t \sim 13.2 \Gyr$. These results make the important point that dynamical friction on a perturber acts to modify the DF of the host (see discussion in Section~\ref{sec:LBKtorque}) , which can introduce features in the DF that cause stalling and/or buoyancy. Hence, predicting a priori whether a perturber will stall or not is difficult without some knowledge of how its inward journey will affect the DF. 

We have seen above that there are two different ways for a perturber to experience stalling; either it stalls at a pre-existing plateau in the DF, as in the case of the \Core halo, or it {\it creates} a plateau in the DF due to how its inspiral exchanges energy and momentum with the host system. An example of the latter is the cNFW halo. Note that these two different scenarios are associated with different manifestations. In the latter case, the creation of the plateau in the DF is associated with a sudden change in the enclosed mass profile of the host halo, evident in Fig.~\ref{fig:cNFWenclosed}. This coincides with the tidal shredding of the core due to the perturber, which happens when the mass enclosed by the perturber becomes comparable to the perturber mass itself \citep[see Fig.~\ref{fig:cNFWdensprof} and][]{Goerdt.etal.10}. In the case of the \Core system, the perturber stalls at a pre-existing plateau. At this stalling radius, the enclosed mass is still much larger than the perturber mass, such that the latter cannot tidally disrupt the core. However, when the mass of the perturber is $M_\rmp=0.01$, it is massive enough to modify the local DF, which can reignite a continued infall, albeit at a very slow rate (see top-right panel of Fig.~\ref{fig:COREenclosed}). Because this subsequent inspiral is so gentle, the tidal distortions of the core remain adiabatic at all times, which explains why no strong distortions of the host system occur. The fact that there are these two different paths to core stalling might explain why in some cases core stalling is found to occur at the radius where the enclosed mass is equal to the perturber mass \citep[see e.g.,][]{Petts.etal.15, DuttaChowdhury.etal.19, DiCintio.Marcos.25}, while in other cases it clearly does not \citep[see][for a detailed discussion]{Dattathri.etal.25b}.

\subsection{Dynamical friction as a catalyst of core collapse}
\label{sec:catalyst}

As shown in Section~\ref{sec:stalling}, when a massive perturber sinks toward the center of its SIDM host halo, it causes drastically accelerated core collapse. For example, the core collapse time of the cNFW halo in the absence of a massive perturber is $t_{\rm cc} \simeq 400 t_0$ (cf. Fig.~\ref{fig:tNFWevol}) which amounts to $\sim 255\Gyr$ ($30\Gyr$) for a cross section of $\sigma_\rmm = 3\cmg$ ($25 \cmg)$. However, as soon as a perturber of mass $M_\rmp = 0.01$ (that is, one percent of the halo virial mass) reaches the core radius, the core collapses in only $\sim 6\Gyr$ ($1\Gyr$). 

Adiabatically growing a mass at the center will increase both the central density and the velocity dispersion gradient. The conductive heat flux is proportional to the product of the conductivity and the gradient of the velocity dispersion (which acts a the effective temperature). In the LMFP regime, the conductivity $\kappa_{\rm LMFP} \propto \sigma_v^3/\lambda_{\rm mfp} \propto \rho \, \sigma_\rmm \, \sigma_v^3$. Hence, as the density and velocity dispersion increase, the conductivity increases as well. Together with the increase in the gradient of the velocity dispersion, this implies a large increase in the conductive flux $\vect{F}_{\rm cond} = \kappa (\rmd\sigma_v^2/\rmd r)$. This explains why adiabatic contraction causes a dramatic speed-up of the core collapse, as already shown by several studies \citep[][]{Elbert.etal.18, Sameie.etal.18, Zhong.etal.23}.

The situation studied here is significantly more complicated. Here, instead of growing a mass at the center, it slowly sinks in via dynamical friction. As a consequence, the orbital energy and momentum lost by the perturber heats the core. Because of the negative heat capacity of gravitational systems, just injecting heat would result in an expansion and cooling of the core, which would delay core collapse, both because of the reduced conductive flux, and because there is more heat to be transferred outward. However, because of the additional increase in core mass, due to the sinking-in of the perturber, the core cannot expand and cool. Rather, it contracts and heats up. This boosts the conductive flux over and above what it would be if the core was merely undergoing core collapse by itself. Note that $\kappa \propto \rho \sigma_v^3$, whereas the heat is proportional to $\sigma_v^2$. Hence, by heating the core without allowing it to expand, one actually speeds up core collapse. As a consequence, transporting a mass inward via dynamical friction is actually more efficient in accelerating core collapse than adiabatically growing the mass at the center. In fact, it is so efficient that the core actually cools as it undergoes core collapse. This is evident from the middle-bottom panel of Fig.~\ref{fig:cNFWdensprof}, which shows that the central velocity dispersion of the host system decreases as the perturber sinks in while the system undergoes core collapse.

\begin{figure}
\centering
\includegraphics[width=0.48\textwidth]{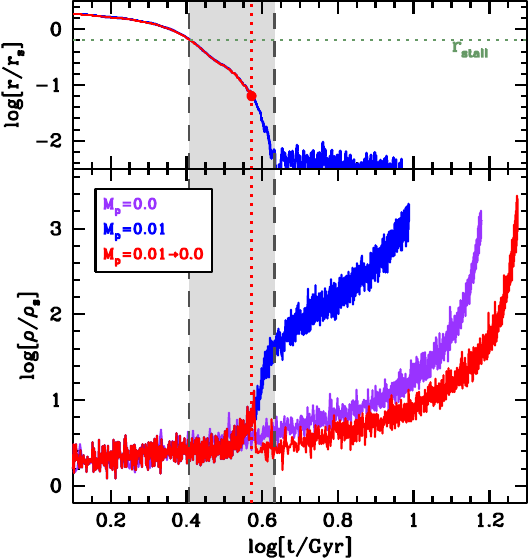}
  \caption{The bottom panel shows the evolution of the central density of the \Core halo in SIDM simulations with $\sigma_\rmm = 10\cmg$. The violet curve shows how the system undergoes core collapse in $\sim 15\Gyr$ in the absence of any perturber. The blue curve shows the results in the presence of a massive perturber of mass $M_\rmp = 0.01$, which at $t=0$ starts out on a circular orbit at $r=2.5 r_\rms$. The top panel shows how the perturber sinks in due to dynamical friction. The gray-shaded region marks the period between the time when the perturber crosses the stalling radius $r_{\rm stall}$, where the perturber {\it would} have stalled in the case of CDM (see Fig.~\ref{fig:core_dynfric}), and the time when it reaches the center of the halo. Note that during this period the core collapse is accelerated compared to the case without perturber. Finally, the red curves show what happens if the perturber is intantaneously removed from the system at $t=3.7 \Gyr$. Its removal causes the core to expand, after which core collapse continues. This illustrates that core collapse is only accelerated while the perturber is sinking through it.}
\label{fig:cendens}
\end{figure}

The bottom panel of Fig.~\ref{fig:cendens} compares the evolution of the central density of the SIDM \Core halo with $\sigma_\rmm = 10 \cmg$ in the absence of a perturber (violet curve) and in the case with $\Mp=0.01$ (blue curve). The top panel shows the evolution of the halo-centric distance of the perturber as it sinks towards the center. The gray shaded region indicates the period between the perturber crossing the stalling radius (where stalling occurs in the case of CDM) and the perturber reaching the center. Note how the perturber accelerates core collapse during this period. Once it has reached the center, core collapse continues, but with a rate that is reminiscent of core collapse in the absence of a perturber. The red curve shows what happens when the perturber is instantaneously removed from the simulation at $t=3.7\Gyr$, when it has reached $r \simeq 0.06 r_\rms$. As is apparent, at that point in time it has already started to accelerate the collapse of the core. However, upon its instantaneous removal, the system reacts by expanding to the impulsive change in the potential, and the central density drops down again. Note that, at the moment of its removal, the halo mass enclosed by the perturber is only approximately ten percent of the mass of the perturber itself, which explains the large response in the halo's density profile. Following the removal of the perturber, the system continues its core collapse, which is now somewhat {\it delayed} compared to the case without a perturber. These results show that {\it core collapse is only accelerated while the perturber sinks through it}. During this period, the core collapse is aided by the gravitational contraction induced by the sinking perturber. When the perturber is removed, the acceleration ceases immediately, indicating that the acceleration is catalyzed, rather than triggered, by the sinking-in of the perturber.

\section{Conclusions}
\label{sec:concl}

Over the years, it has become clear that gravitational $N$-body systems with a core, roughly defined as a central region with a logarithmic density gradient $-0.5 \lta \rmd\log\rho/\rmd\log r \leq 0$, are subject to intriguing dynamical phenomena. As first predicted by \citet{Weinberg.86}, and later demonstrated by \citet{Read.etal.06c}, using numerical simulations, dynamical friction can cease to operate inside cores, leading to core stalling. In later work, \citet{Cole.etal.12} showed, again using simulations, that a massive object positioned within such a core experiences a dynamical buoyancy that causes it to move outward. Finally, \citet{Dattathri.etal.25a} demonstrated that cores can be unstable to a rotating dipole mode.

Recently, \citetalias{Dattathri.etal.25b} presented a unified picture of these unexpected dynamical phenomena grounded in kinetic theory. In particular, core stalling, dynamical buoyancy, and the dipole instability are all shown to be related to specific features (plateaus and inflections) in the phase-space DF. Using high resolution numerical simulations they showed that the sign of the torque on the perturber is set by the sign of $\rmd f/\rmd E$ at the orbital energy of the perturber, here called $(\nabla f)_{\rm CR}$.  Dynamical friction and buoyancy correspond to $(\nabla f)_{\rm CR} < 0$ and $>0$, respectively, while a vanishing $(\nabla f)_{\rm CR}$ gives rise to core stalling. This is exactly what is predicted based on the expression for the LBK torque of equation~(\ref{LBK1}) as long as corotation is the dominant resonance. 

An important corollary is that, since the DF evolves as the perturber sinks inward, it is difficult to predict a priori whether or not a perturber will stall in a certain system. For example, a massive perturber may start to sink inside a system that has $\rmd f/\rmd E<0$ everywhere, but modify the DF as it sinks in to such an extent that it develops an inflection, causing subsequent stalling and or buoyancy. Similarly, a perturber may stall at a plateau or bump where $(\nabla f)_{\rm CR} \simeq 0$, modify the DF by nonlinear effects such as particle trapping, and then re-ignite friction (or buoyancy) when the local gradient in the DF has become nonzero. Explicit examples of such behavior are shown in Figs.~\ref{fig:cNFW_dynfric},~\ref{fig:coreMsat}, and~\ref{fig:core_dynfric} and in \citetalias{Dattathri.etal.25b}.

All previous studies of core dynamics have focused on collisionless systems, either galaxies or CDM halos. In this paper we studied whether and how core dynamics differ in the case of SIDM, which is known to produce cored dark matter halos. Our main conclusions are:
\begin{itemize}[leftmargin=0.27truecm, labelwidth=0.2truecm]
    
  \item{Self-interactions act to suppress core stalling, dynamical buoyancy, and the dipole instability. The reason is that self-interactions cause a diffusion in phase space that acts to erase small-scale features in the DF and drive the system's DF toward a featureless exponential, $f(E) \propto \exp(-E/\sigma^2)$, characteristic of an isothermal system. Since the corresponding $\rmd f/\rmd E$ is always negative, cores in SIDM halos are not expected to be subject to any of the core dynamical processes discussed here. Obviously, the extent to which self-interactions can suppress stalling and buoyancy depends on the cross section, with larger cross sections resulting in stronger and faster suppression}.

  \item{As the massive perturber sinks inside the core of an SIDM halo, it catalyzes core collapse by gravitationally contracting the core. This can be extremely efficient, shortening the time to reach core collapse by over an order in magnitude.}
    
\end{itemize}

\noindent
These findings have a number of important implications. Most importantly, the fact that the dynamics in CDM and SIDM cores can be very different opens a potential avenue for constraining the nature of dark matter. Dipole instabilities, buoyancy, and stalling have a number of observational consequences. Among others, they give rise to lopesidedness in the central regions of galaxies and to nuclear star clusters and AGN that are offset from the galaxy's center inferred from the outer isophotes. Interestingly, such offsets are not uncommon \citep[e.g.,][]{Bingelli.etal.00, Cote.etal.06, Shen.etal.19, Reines.etal.20, Sargent.etal.22, Mezcua.etal.24}. Taken at face value, this seems to favor CDM over SIDM. However, it is important to account for the fact that there are various other explanations for offset NSCs or AGN, including, but not restricted to, recent accretion (time since accretion is less than the dynamical friction time scale), galaxy mergers, recoil kicks following a SMBH merger \citep[][]{Campanelli.etal.07}, or dynamical heating by substructure \citep[][]{Boldrini.etal.20}. Nevertheless, we are optimistic that future studies with better statistics, based on larger data sets and simulation suites, will be able to put meaningful constraints on the self-interaction cross sections for the dark matter.

Related to this, core stalling and buoyancy can potentially delay or prevent the merger of two or more SMBHs, which has implications for the total gravitational wave signal to be measured by LISA and/or pulsar timing arrays, and may also impede the merging of globular clusters in dwarf galaxies. The latter is one of the leading scenarios for the formation of nuclear star clusters \citep[][]{Tremaine.etal.75, CapuzzoDolcetta.Miochi.08, ArcaSedda.CapuzzoDolcetta.16}. Our results suggest that these processes will be more efficient in SIDM than in CDM in the case where both produce cores of similar densities.

Using cosmological simulations, \citet{DiCintio.etal.17} find that in low mass galaxies (halo masses of $\sim 10^{10.5}-10^{10.9} \Msun$), about 80 percent of SMBHs are off-centered in the case of SIDM, compared to only $\sim 10$ percent in the case of CDM. The authors ascribe this difference to core stalling operating in the SIDM halos. At first sight, this seems opposite to our `prediction'. However, whereas their SIDM halos develop a central core due to self-interactions, the CDM halos maintain an NFW-like central cusp. Hence, the SIDM and CDM halos do not have similar central densities. Furthermore, their simulations do not properly resolve the dynamical friction acting on the SMBHs. Instead, they adopt a dynamical friction subgrid model developed by \citet{Tremmel.etal.15} that closely follows the Chandrasekhar prescription. Since this models the dynamical friction force as being directly proportional to the local density, dynamical friction is more efficient in the CDM halos where the central densities are higher. However, this methodology does not adequately account for the intricate dynamical processes highlighted here, which ultimately arise from resonance interactions. Higher resolution simulations that properly resolve resonance interactions between the SMBHs and the dark matter or star particles in the central regions are required to properly resolve core stalling and buoyancy effects. Until then, it remains to be seen whether or not, in a fully self-consistent, cosmological setting, SIDM cores indeed produce less stalling than CDM cores of similar density, as argued here.

Our findings also impact the conclusions of several studies that invoked or relied on core stalling. In their review paper, \citet{Adhikari.etal.25} argue that, because of core stalling, one ought to expect a lack of dynamical friction in SIDM cores. However, the results presented here argue that this is not the case and that CDM and SIDM cores can display very different dynamics. \citet{Modak.etal.23} argued that the formation of nuclear star clusters via the merging of globular clusters requires the presence of a cusp, since cores cause core stalling. In light of the results presented here, this conclusion is subject to the assumption that dark matter self-interactions are negligible. Similarly, core stalling and buoyancy have been invoked as a solution to the timing problem of globular clusters in dwarf spheroidals \citep[][]{Goerdt.etal.06, Cole.etal.12, Boldrini.etal.19, SanchezSalcedo.Lora.22}, or the presence of multiple nuclei in the centers of galaxies \citep[]{Bonfini.Graham.16, Meadows.etal.20, Nasim.etal.21}. As shown here, these results depend on the nature of dark matter; in the case of SIDM one can only invoke these processes to solve the timing problem if the self-interaction cross sections are sufficiently small. Hence, future studies tailored to individual systems such as the Fornax galaxy, which is one of the galaxies for which the timing problem is the most severe, again have the potential to constrain the SIDM self-interaction cross section.

Finally, the fact that massive perturbers that sink inside SIDM cores catalyze core collapse may have important ramifications. For example, \citet{Slone.etal.23} argue that the observed central densities of Milky Way satellite galaxies such as Draco and Ursa Minor can be used to constrain the self-interaction cross sections of SIDM models. In particular, they argue that one can rule out a significant part of SIDM parameter space as it predicts that the dark matter haloes of the dwarf spheroidals in question would have produced overly large cores that are inconsistent with their high observed central densities. However, this did not account for the fact that the accretion of a massive object such as a supermassive black hole during the dwarf spheroidal's hierarchical formation history might have drastically shortened the core collapse time. Hence, it is crucial to first rule out that the nuclei of the dwarf spheroidals used for this test are devoid of any central massive objects.

In closing, we emphasize that all simulations used here are highly idealized. In particular, we have restricted ourselves to spherical, isotropic host halos and to perturbers that are modeled as solid bodies with a Plummer potential moving along circular orbits. In reality, halos are triaxial, anisotropic, and evolving in time, and perturbers are typically subject to tidal stripping and heating while moving along eccentric orbits. How each of these complications impact the results presented here is left for future work.


\section*{Acknowledgements}

In loving memory of Avishai Dekel, whose kindness, encouragement, and physical insight have been a source of inspiration throughout our careers. We are grateful to Uddipan Banik, Kimberly Body, Annika Peter, Oren Slone, Charlie Mace, Fangzhou Jiang, and Martin Weinberg for useful discussions, and to Anne Lisa Vari, Jean-Baptiste Fouvry, Matthew Kunz, and Jonathan Squire for organizing the workshop ``Interconnections between the Physics of Plasmas and Self-gravitating Systems" at the KITP in Santa Barbara where part of this work was performed. The KITP Santa Barbara is supported in part by the National Science Foundation (NSF) under Grant No. NSF PHY-2309135. This work has been supported by the NSF through grant AST 2407063.


\bibliographystyle{mnras}
\bibliography{references_vdb} 


\appendix
\numberwithin{figure}{section}
\numberwithin{table}{section}
\numberwithin{equation}{section}

\section{Gravothermal Fluid Equations}
\label{sec:gravothermal}

Consider a spherical SIDM halo with a density profile $\rho(r,t)$ and an enclosed mass $M(r,t)$ at radius $r$ and time $t$. Assume the halo to be isotropic and in quasi-static virial equilibrium with a one-dimensional velocity dispersion $v(r,t)$. Due to self-interactions, characterized by a cross section per unit mass $\sigma_\rmm$, heat flows in the radial direction with luminosity $L(r,t)$ through a spherical shell at radius $r$. The corresponding heat flux obeys Fourier's law of thermal conduction, 
\begin{equation}\label{thermalconduct}
\frac{L}{4 \pi r^2} = -\kappa \frac{\partial T}{\partial r}\,,
\end{equation}
with $\kappa$ conductivity and $k_\rmB T = m\,v^2$ a measure of the temperature of the particles of mass $m$. The temporal evolution of such a halo is a time-dependent diffusion problem that can be modeled as a gravothermal fluid \citep[][]{LyndenBell.Eggleton.80, Balberg.etal.02}. In addition to the heat conduction equation, such a fluid obeys mass conservation,
\begin{equation}\label{massconservation}
\frac{\partial M}{\partial r} = 4 \pi r^2 \rho\,,
\end{equation}
hydrostatic equilibrium, 
\begin{equation}\label{hydrostatic}
\frac{\partial \left(\rho v^2\right)}{\partial r} = -\frac{G M \rho}{r^2}\,,
\end{equation}
where we have defined the pressure $P = \rho v^2$, and the first law of thermodynamics which can be written as
\begin{equation}\label{firstlaw}
\frac{\partial L}{\partial r} = -4 \pi \rho r^2 v^2 \left( \frac{\partial}{\partial t} \right)_M \ln(v^3/\rho)\,,
\end{equation}
with $s = \ln(v^3/\rho)$ an effective entropy. Each time step, heat flows in the radial direction as dictated by equation~(\ref{thermalconduct}), after which the system re-establishes hydrostatic equilibrium, while satisfying mass conservation and the first law of thermodynamics. 

Strictly speaking, Fourier's law of thermal conduction is valid in the short mean-free path (SMFP) regime, where the Knudsen number $\Kn \ll 1$. Here $\Kn \equiv \lambda/H$ with $\lambda = (\rho \sigma_\rmm)^{-1}$ the mean-free path and $H =\sqrt{v^2/(4 \pi G \rho)}$ the gravitational Jeans length, which is a measure for the size of a virialized system. In the SMFP regime, Chapman-Enskog theory \citep[][]{Lifshitz.Pitaevskii.81} dictates that the conductivity is given by
\begin{equation}\label{condSMFP}
\kappa_{\rm SMFP} = \frac{3}{2} a^{-1} b \, n \, k_\rmB \, {\lambda^2 \over t_\rmr} 
\,,
\end{equation}
\citep[][]{Balberg.etal.02, Koda.Shapiro.11}, where $n=\rho/m$ is the number density of particles, $k_\rmB$ is the Boltzmann constant, $b=25\sqrt{\pi}/32$ is the effective impact parameter, and $t_\rmr \equiv \lambda/(av)$ is the collision time (cf. equation~[\ref{tcoll}] in the main text) with $a = 4/\sqrt{\pi}$ a coefficient relevant for hard-sphere scattering of particles with a Maxwell-Boltzmann distribution \citep[][]{Reif.65}. In the LMFP regime, which is more common for SIDM halos, no formal expression for the conductivity exists, but \citet{LyndenBell.Eggleton.80} have shown that an empirical relation of the form
\begin{equation}\label{condLMFP}
\kappa_{\rm LMFP} = \frac{3}{2} C \, n \, k_\rmB \, {H^2 \over t_\rmr}\,, 
\end{equation}
with $C$ a free calibration parameter of order unity (see below), can adequately describe the gravothermal collapse of globular clusters. Following \citet{Balberg.etal.02}, we interpolate between these two regimes using $\kappa^{-1} = \kappa_{\rm SMFP}^{-1} + \kappa_{\rm LMFP}^{-1}$.

Given appropriate boundary conditions, equations~(\ref{thermalconduct})-(\ref{firstlaw}) can be solved given a set of initial radial profiles. Typically, these are the profiles of a virialized, collisionless system, which is also the approach used to set up the initial conditions for the $N$-body simulations\footnote{This approach is valid as long as the impact of the SIDM self-interactions is small during the process of halo formation via violent relaxation.}. We segment the halo into $N=200$ concentric shells, logarithmically spaced between $r_{\rm min} = 0.01 r_\rms$ and $r_{\rm max} = 100 r_\rms$, and closely follow the methodology of \citet{Nishikawa.etal.20} to evolve the system in time. 

The conduction parameter $C$, which controls the efficiency of heat transfer in the LMFP regime, needs to be calibrated against numerical simulations. We do so by comparing the evolution of the central density of the tNFW halo in the numerical simulation to that obtained using the gravothermal fluid equations. This yields $C=0.76$ (red-dashed line in top-left panel of Fig.~\ref{fig:tNFWevol}), which is in excellent agreement with \citet{Koda.Shapiro.11} and \citet{Nishikawa.etal.20}, who obtained a similar value for $C$ to match the evolution of an untruncated NFW profile. Using $C=0.76$ also accurately fits the evolution of the \Core halo (red-dashed line in bottom-left panel of Fig.~\ref{fig:tNFWevol}. For completeness, we point out that some studies have advocated different calibration values for $C$. In particular, \citet{Essig.etal.19} find $C\sim 0.6$, while a recent study by \citet{Mace.etal.25} advocates for $C\sim 1.0$\footnote{Both \citet{Essig.etal.19} and \citet{Mace.etal.25} refer to $C$ as $\beta$.}. The reason for these discrepancies is currently unclear. 

\section{Calculating the phase-space distribution function}
\label{sec:compDF}

In order to compute the evolution of the DF in our simulations, we proceed as follows. Given the phase-space coordinates of all particles in a simulation output, we compute for each particle its energy $E = \frac{1}{2} v^2 + \Phi(\vec{r})$ and angular momentum $L = \vert \vec{r} \times \vec{v} \vert$. These are used to infer
\begin{equation}\label{dNdEdL}
   \frac{\rmd^2 N(E,L)}{\rmd E\,\rmd L} = f(E,L) \, g(E,L)\,,
\end{equation}
where $f(E,L)$ is the DF, and 
\begin{equation}\label{dens_states}
    g(E,L) = 16 \pi^2 L \int_{r_\rmp}^{r_\rma} \frac{\rmd r}{\sqrt{2(E-\Phi)-(L/r)^2}} \,,
\end{equation}
is the density of states. Here $r_\rmp$ and $r_\rma$ are the pericenter and apocenter distances for an orbit with energy $E$ and angular momentum $L$, which, in a spherical potential, are the roots of
\begin{equation}
 \frac{1}{r^2} + \frac{2(\Phi-E)}{L^2} = 0\,.
\end{equation}

We assume that the system remains spherical throughout its evolution, so that the density of states follows from equation~(\ref{dens_states}) and the instantaneous spherically averaged potential. The instantaneous DF then follows from equation~(\ref{dNdEdL}). 

Note that while we start out with an isotropic DF, $f=f(E)$, the sinking of the perturber transfers momentum to the halo particles, which may cause the DF to develop a dependence on $L$. Hence, in computing the DF, we consider the more generic form $f=f(E,L)$. In practice, though, we typically infer the DF to have only a weak $L$-dependence. Therefore, in what follows, we compute the one-dimensional DF $f(E)$, which we obtain by marginalizing $f(E,L)$ over $L$, that is,
\begin{align}\label{eq:marginalization}
f(E) = \dfrac{\int_0^{L_\rmc\left(E\right)} \rmd L\, f(E,L) g(E,L) }{\int_0^{L_\rmc\left(E\right)} \rmd L\, g(E,L)}\,,
\end{align}
with $L_\rmc(E)$ the angular momentum of a circular orbit of energy $E$. For isotropic systems, the $f(E)$ thus obtained is identical to that inferred from its density distribution using the Eddington inversion.

\section{The Knudsen Number}
\label{sec:Knudsen}

\begin{figure*}
\centering
\includegraphics[width=0.9\textwidth]{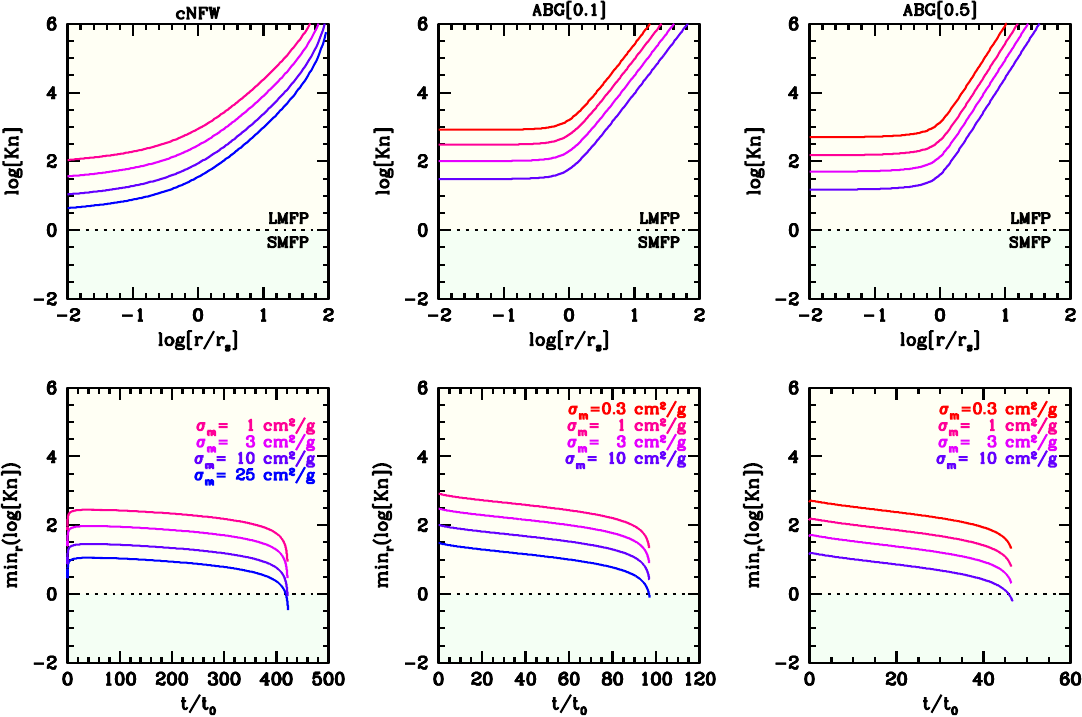}
  \caption{{\it Top panels:} The value of the Knudsen number, $\Kn$, in the initial conditions as a function of radius for all three halos discussed in the paper (different columns) and for all different values of the collisional cross section used (different colors), as indicated. The dashed vertical lines corresponds to $\Kn=1$, which separates the small mean-free-path (SMFP) regime ($\Kn<1$, indicated in light-green) from the LMFP regime ($\Kn>1$, indicated in light-yellow). {\it Bottom Panels:} temporal evolution, computed using the gravothermal fluid equations, of the mimimum value of the Knudsen number }
\label{fig:Knudsen}
\end{figure*}

Fig.~\ref{fig:Knudsen} shows information regarding the Knudsen number (see Section~\ref{sec:sims}) in our simulations. Top panels show the radial profile of the Knudsen number in the initial conditions, while the bottom panels show the temporal evolution of the minimum Knudsen number (typically that at the center). The latter is obtained using the gravothermal fluid model of Appendix~\ref{sec:gravothermal}; the results from the $N$-body simulations are similar but far more noisy. The three different columns correspond to the three different halos considered in this study, while lines of different colors correspond to different values for the collisional cross section, as indicated in the lower panels. Note that all our simulations are, at all times, in the LMFP regime (shaded in yellow), with the exception of the simulations with the largest cross sections, for which the minimum Knudsen number becomes slightly smaller than unity during the final stages of core collapse modeled here.

\vfill
\eject


\label{lastpage}

\end{document}